\documentclass[12pt,aps,prb,preprint]{revtex4}   % style for Physical Review B and AJP are similar

\usepackage{amsmath}    % need for subequations
\usepackage{amsfonts}   % note how statements can be commented out
\usepackage{amssymb}
\usepackage{graphicx}   % for figures
\begin{document}

\title{McMillan-Mayer Theory of Solutions Revisited: Simplifications and Extensions}
\author{Shaghayegh Vafaei}
 \email{svafaei@uoguelph.ca}   %optional
\affiliation{Department of Physics and Biophysics Interdepartmental Group, University of Guelph, Guelph, ON, N1G2W1, Canada}
\author{Bruno Tomberli}
 \email{brunotomberli@capilanou.ca} 
\affiliation{Department of Physics, Capilano University, Vancouver, BC, V7J3H5, Canada}
\author{C.G.Gray}
 \email{cgray@uoguelph.ca} 
\affiliation{Department of Physics, University of Guelph, Guelph, ON, N1G2W1, Canada}
\date{\today}

\begin{abstract}
McMillan and Mayer (MM) proved two remarkable theorems in their paper on the equilibrium statistical mechanics of liquid solutions. They first showed that the grand canonical partition function for a solution can be reduced to a one with an effectively solute-only form, by integrating out the solvent degrees of freedom. The total effective solute potential in the effective solute grand partition function can be decomposed into components which are potentials of mean force for isolated groups of one, two, three, etc, solute molecules. Secondly, from the first result, now assuming low solute concentration, MM derived an expansion for the osmotic pressure in powers of the solute concentration, in complete analogy with the virial expansion of gas pressure  in powers of the density at low density. The molecular expressions found for the osmotic virial coefficients have exactly the same form as the corresponding gas virial coefficients, with potentials of mean force replacing vacuum potentials. In this paper we restrict ourselves to binary liquid solutions with solute species $A$ and solvent species $B$  and do three things: (a) By working with a semi-grand canonical ensemble (grand with respect to solvent only) instead of the grand canonical ensemble used by MM, and avoiding graphical methods, we have greatly simplified the derivation of the first MM result, (b) by using a simple nongraphical method developed by van Kampen for gases, we have greatly simplified the derivation of the second MM result, i.e., the osmotic pressure virial expansion; as a by-product, we show the precise relation between MM theory and  Widom potential distribution theory,  and (c) we have extended MM theory by deriving virial expansions for other solution properties such as the enthalpy of mixing. The latter expansion is proving useful in analyzing ongoing ITC experiments with which we are involved. For the enthalpy virial expansion  we have also changed independent variables from semi-grand grand canonical, i.e., fixed $\{N_A,\mu_B,V,T\}$, to those relevant to  the experiment, i.e.,  fixed $\{N_A,N_B,p,T\}$,  where $\mu$ denotes chemical potential, $N$ the number of  molecules, $V$ the volume, $p$ the pressure, and $T$ the temperature.

\end{abstract}

\maketitle

\section{INTRODUCTION}

In a classic paper\cite{mcmillan-mayer} on the  statistical mechanical theory of solutions in equilibrium McMillan and Mayer (hereafter MM) derived the following two results (or theorems):
 
 {\bf MMI}: For a solution with any number of components, the solvent variables in the grand canonical partition function can be formally ``integrated out'' leaving an effective partition function for the solutes only. The total effective solute potential entering the effective solute partition function is the true vacuum solute interaction potential plus a contribution induced by the integrated-out solvent. The total effective solute potential can be decomposed into a sum of one-body, two-body, etc, effective solute potentials.  These  effective solute potentials are potentials of mean force (PMFs) for the solute molecules.
 
 {\bf MMII}: Using the results of MMI, the osmotic pressure  $\pi$ of a dilute solution can be developed in a virial series in the solute concentration, analogous to the gas virial expansion for the pressure $p$ in powers of the gas  density at low density. The statistical mechanical expressions for the osmotic pressure virial coefficients have exactly the same form as the expressions for the gas pressure virial coefficients, with solute PMFs replacing gas vacuum potentials. 
 
A few years after the MM paper appeared, another classic paper \cite{kirkwood-buff} on the statistical mechanical  theory of solutions in equilibrium was published by Kirkwood and Buff (hereafter KB). This paper gives molecular expressions for specific solution macroscopic properties, i.e.,  the isothermal compressibility, the partial molecular volumes, and the inverse osmotic susceptibilities,  in terms of what are now called KB integrals. The latter are the integrals $\int d \boldsymbol{r}(g_{\alpha \beta}(r)-1)$ over the solution pair correlation functions $g_{\alpha \beta}(r)$, where $\alpha$ and $\beta$ label the molecular species, e.g., $A$ and $B$ for a two-component solution, and $r$ is the separation between the centers of an $\alpha \beta$ pair of molecules.  The pair correlation functions themselves can in principle be determined \cite{hansen-mcdonald, gray-gubbins, gray-gubbins-joslin} both theoretically, from liquid state simulations and analytic methods using model intermolecular potentials, and  experimentally, from X-ray and neutron diffraction experiments. KB theory has been simplified, extended, reviewed extensively \cite{ben-naim, gray-gubbins-joslin, ploetz-smith, matteoli-mansoori}, and widely applied \cite{blanco-et-al, liu-ruckenstein, barbosa-widom, tomberli-et-al, matteoli-mansoori, villa-et-al}. One leading practitioner has expressed the opinion ``I believe that in all regards the KB theory is immensely superior to the MM theory, as I hope to convince the reader...''. We hope to convince the reader that, with current and future developments, MM theory could be far more useful than is now the case, although there is no doubt that at the present time, particularly since the introduction of the Ben Naim KB inversion technique\cite{ben-naim1977, ben-naim, gray-gubbins-joslin,ploetz-smith, matteoli-mansoori}, which expresses the KB  integrals in terms of the specific  macroscopic properties mentioned above, KB theory is used much more.

In contrast to KB theory, MM theory has not been greatly simplified or extended, and has been reviewed \cite{hill1960, ben-naim} infrequently. It also has not been  widely applied, with one exception: the osmotic pressure virial expansion. For a two-component solution with dilute solute (species A) in a solvent (species B), the MM result for the osmotic pressure $\pi$ as a virial series in solute number density $c_A = N_A/V$, where $N_A$ is the number of solute molecules and $V$ is the system volume, is 

%====================================================
% 1.1
\begin{equation}
\label{eq1.1}
\pi = c_Ak_BT[1 + \widetilde{B}(\mu_B,T)c_A +...]~ ,
\end{equation} 
%====================================================
where $k_B$ is Boltzmann's constant, $T$ the temperature,~ and $\mu_B$ the solvent chemical potential. The first virial term $c_Ak_BT$ is the famous van't Hoff solute one-body term which is universal, i.e., independent of all intermolecular interactions, and has the same form as the ideal gas pressure. Solution properties whose first virial term depends only on the solute  number density and is independent of solute molecular details and interactions  are called colligative. The second term, with coefficient  $\widetilde{B}$, is $O(c_A^2)$ and takes into account the contribution from pairs of interacting solute molecules in the solvent, and the MM  molecular expression for this  osmotic second virial coefficient is

%====================================================
% 1.2
\begin{equation}
\label{eq1.2}
\widetilde{B}(\mu_B,T) = -\frac{1}{2}\int d\boldsymbol{r}( \mbox{e}^{-\beta w_{AA}(r)} - 1 )~  ,
\end{equation}
%=======================================================
where $\beta = 1/k_BT $, and   $w_{AA}(r)$ is the PMF for a pair of solute molecules at infinite dilution in the solvent, which is written as an explicit function of solute-solute separation $r$ and which also depends implictly on solvent chemical potential and temperature. Note that (\ref{eq1.2}) has the same form as the  expression for the gas second virial coefficient,  with pair PMF $w_{AA}(r)$ in place of the vacuum pair potential $u_{AA}(r)$ which occurs in the gas second virial coefficient expression. This is an example of MMII above.
Equation (\ref{eq1.1}) has long been used \cite{flory, doi, tanaka, steiner-garone, moon-et-al, velev-et-al, asthagiri-et-al, tavares-et-al, siderius-et-al, blanco-et-al, lima-et-al}   by polymer, biopolymer and colloid solution researchers to obtain molecular weights of macromolecules (from the first virial term $c_Ak_BT$) and to study macromolecule solute-solute interactions (using the second virial term). The osmotic second virial coefficient has also been used in discussions of the hydrophobic interaction of small nonpolar solutes in water\cite{rossky-friedman,wood-thompson, ludemann-et-al}, and we give an example in the concluding section.

Unfortunately the MM paper \cite{mcmillan-mayer} is very hard to read and this is likely the main reason for the relative neglect of MM theory compared to KB theory. Indeed, as another leading practitioner \cite{widom2011} has put it ``...somewhere, hidden in an impenetrable jungle of notation in the 1945 paper of McMillan and Mayer, must be the potential distribution theorem!''  As it turns out the Widom potential distribution theorem\cite{widom1963} (also called the particle insertion theorem) is not explicitly written down by MM (Hill\cite{hill1960} comes close in his reformulation of MM theory using the semi-grand canonical ensemble-see remarks following). We find that the Widom expression for the excess chemical potential of a solute molecule in an otherwise pure solvent is the first virial term in the virial expansion of the solute  chemical potential $\mu_A$; the latter expansion is obtained  by differentiating the semi-grand potential virial series with respect to solute number $N_A$. The semi-grand potential is the thermodynamic potential generated by the semi-grand canonical ensemble (grand with respect to solvent only), the ensemble we employ in this paper. MM did not obtain the semi-grand potential series because they worked with the full grand canonical ensemble which generates the grand potential. Because the grand potential is proportional to the pressure, MM obtained the virial series (\ref{eq1.1}) for the osmotic pressure directly, and because this series has a trivial first virial term, a nontrivial first virial term which occurs for non-colligative properties such as the solute chemical potential is not visible explicitly. 

 Some of the difficulty of the MM paper has been reduced by Hill \cite{hill1960} who separates better the quantum and classical discussions, does not include solid solutions in the discussion,  discusses explicitly a two-component classical liquid solution, and introduces a semi-grand canonical ensemble to derive MMI.  In this paper we greatly simplify the presentation  further by working with the semi-grand canonical ensemble in a new way to derive MMI as described in the next paragraph, and by using a simple and nongraphical virial expansion method \cite{vankampen} to derive MMII from MMI. We also extend MM theory by deriving virial  expansions for other properties such as the enthalpy of mixing, which are needed to analyze recent isothermal titration calorimetry (ITC) experiments \cite{nichols-et-al} on dilute  peptide aqueous solutions.

In the next section, for a classical  two-component liquid solution we introduce a semi-grand canonical ensemble which is grand canonical with respect to the solvent(B) and canonical with respect to the solute(A); the complete set of independent variables is $\{N_A,\mu_B,V,T\}$. These state conditions are the natural ones for analysis of osmotic pressure experiments where the solvent and solution are in contact through a semi-permeable membrane (so that $\mu_B$ is fixed) and the solution pressure is measured as a function of the controlled solute number $N_A$ at fixed solution volume $V$ and temperature $T$. Various semi-grand ensembles have been discussed in the literature, starting with Stockmayer \cite{stockmayer} (see also \cite{hill1960, rosgen-et-al, siderius-et-al}) who, in order to discuss light scattering from  multicomponent  dilute solutions,  introduced a constant pressure  ensemble which is grand with respect to the solutes but not with respect to the solvent. The semi-grand canonical ensemble we employ was introduced by Hill \cite{hill1960}. This ensemble was also used to derive MMI by  Kodytek\cite{kodytek}, who, however, needlessly complicated the derivation by introducing the total solute PMF in the canonical ensemble, and by  Dijkstra et al \cite{dijkstra-et-al}. Our discussion of MMI is similar in spirit  to those of Hill\cite{hill1960} and Dijkstra et al\cite{dijkstra-et-al}, but simpler, because rather than focussing on the  semi-grand potential for the solution,  we focus instead on the difference between the solution and solvent semi-grand potentials, which can be expressed in terms of the logarithm of the  ratio of the solution and solvent semi-grand partition functions. As with other such problems in statistical mechanics involving the difference in free energies for two systems with some state conditions in common,  such as  Zwanzig's thermodynamic perturbation theory\cite{zwanzig, gray-gubbins,raineri-et-al, Lu-et-al}, van Kampen's gas virial expansion method\cite{vankampen}, and Widom's particle insertion theorem\cite{widom1963,gray-gubbins-joslin}, the ratio of partition functions generates  automatically the required solvent-to-solution  change in the form of a normalized average, and this average can be expressed in terms of the total solute PMF and its one-body, two-body, etc, components  from simple physical considerations. Thus, for example, in contrast to reference \cite{dijkstra-et-al}, in discussing MMI we do not need graphical methods, and a zero-body term in our total solute PMF does not arise. This will become clearer in Section III. A similar semi-grand ensemble but with fixed pressure rather than fixed volume has been introduced for computer simulations of multicomponent solutions \cite{kofke, morrow-maginn}.

In Section III we introduce the solute PMFs, first $W_A(\boldsymbol{r}_A^{N_A})$ for the complete  set of $N_A$ solute molecules as a function of their configuration $\boldsymbol{r}_A^{N_A}$ , and then  reduced PMFs for isolated groups of one, two, three, etc, solute molecules alone in the solvent, i.e., $w_A(\boldsymbol{r}_{1}), w_{AA}(r_{12}), w_{AAA}(r_{12},r_{13},r_{23})$, etc., respectively, where $\boldsymbol{r}_{1}$ is the position of solute molecule $A_1$, $r_{12}$ the separation of solute molecules $A_1$ and $A_2$, etc. These are related in the usual way to the corresponding solute reduced distribution and correlation functions, and to mean forces on solute molecules under the proper constraints \cite{hill1960}. The semi-grand canonical ensemble partition function introduced in Section II is re-expressed in terms of an effective solute-only canonical ensemble partition function involving $W_A(\boldsymbol{r}_A^{N_A})$. The total solute PMF or effective potential $W_A(\boldsymbol{r}_A^{N_A})$ is then decomposed into a sum of its one-body, two-body, etc, components, involving $w_A(\boldsymbol{r}_{1}), w_{AA}(r_{12})$, etc.  This is the semi-grand canonical ensemble version of MMI. For the reasons discussed, we believe this derivation of MMI to be much simpler than those presented previously.

The virial expansions are introduced in Section IV. Starting with the result of Section III expressing the semi-grand canonical partition function in terms of the effective solute-only canonical ensemble  partition function involving the effective solute potential $W_A(\boldsymbol{r}_A^{N_A})$ and its decomposition into one-body, two-body, etc, components, we assume a dilute solution and derive a virial series in $c_A$ for the difference between the solution and solvent semi-grand potentials, $\widetilde{F} - \widetilde{F}_B$, from which we immediately obtain the virial series for the osmotic pressure $\pi = p - p_B$, where $p$ and $p_B$ are the solution and solvent pressures, both at chemical potential $\mu_B$. This is the semi-grand canonical version of MMII.  Our derivation is much simplified compared to that of MM by using a simple nongraphical canonical ensemble virial expansion method developed for gases \cite{vankampen}. Another advantage, mentioned above, is that the virial series for $\tilde{F}$, unlike that for $\pi$, has a nontrivial first virial coefficient, which is needed for other non-colligative properties such as solute chemical potential and solubility, and we  extend MM theory and derive virial expansions for various such properties including the enthalpy of mixing. For some of  the extensions, we also change independent variables from semi-grand canonical variables $\{N_A,\mu_B,V,T\}$ to those more relevant to the particular experiment, e.g., $\{N_A,N_B,p,T\}$ for the enthalpy of mixing experiment. In this section we also clarify the relation between MM theory and Widom's potential distribution theory.

The paper concludes with a simple example, and some remarks on ongoing work by our group.

%%%%%%%%%%%%%%%%%%%%%%%%%%%%%%%%%%%%%%%%%%%%%%%%%%%%%%%%%%%%%%%%%%%%%%%%%%%%%%%%%%%%%%%%%%%%%%%%%%%%%%%%%%%%%%%%%%%%%%

\section{STATISTICAL MECHANICS IN A SEMI-GRAND CANONICAL ENSEMBLE}

%%%%%%%%%%%%%%%%%%%%%%%%%%%%%%%%%%%%%%%%%%%%%%%%%%%%%%%%%%%%%%%%%%%%%%%%%%%%%%%%%%%%%%%%%%%%%%%%%%%%%%%%%%%%%%%%%%%%%%

We first consider the classical statistical mechanics of a binary liquid solution of $N_A$ identical solute molecules and $N_B$ identical solvent molecules in equilibrium at temperature $T$ in a volume $V$. For notational simplicity we consider monatomic molecules but, as we indicate later, the final results are immediately generalizable to polyatomic molecules which may be nonrigid.  We write the Hamiltonian $H$ as 

%==========================================================
% 2.1
\begin{equation}
\label{eq2.1}
H = \sum_{i=1}^{N_A}\frac{p_{i}^{2}}{2m_A} + \sum_{j=1}^{N_B} \frac{p_{j}^{2}}{2m_B} + U(\boldsymbol{r}_{A}^{N_A}, \boldsymbol{r}_B^{N_B})  , 
\end{equation}
%=========================================================
where $\boldsymbol{p}_i$ denotes the momentum of molecule $i$, $m_A$ the mass of an $A$-molecule, $ \boldsymbol{r}_{A}^{N_A}$ the configuration of the $A$-molecules, and assuming pairwise additive vacuum potentials, the total potential energy $U$ is given by

%=========================================================
% 2.2
\begin{align}
\label{eq2.2}
 U(\boldsymbol{r}_{A}^{N_A}, \boldsymbol{r}_B^{N_B}) &= \sum_{i< k}^{N_A}u_{AA}(r_{ik}) +  \sum_{j < l}^{N_B}u_{BB}(r_{jl}) + \sum_{i=1}^{N_A}\sum_{j=1}^{N_B}u_{AB}(r_{ij}) 
 \nonumber \\ &\equiv U_{AA}(\boldsymbol{r}_{A}^{N_A}) + U_{BB}(\boldsymbol{r}_{B}^{N_B}) + U_{AB}(\boldsymbol{r}_{A}^{N_A} ,\boldsymbol{r}_{B}^{N_B}).                                                    
\end{align}       
%==========================================================
Here $U_{AA}, U_{BB}$ and $U_{AB}$ denote the total solute-solute, solvent-solvent and solute-solvent potentials, respectively, and $r_{ij}$ the separation between molecules $i$ and $j$. The assumption that $U_{AA}$, etc, are pairwise additive is again for notational convenience, and without this assumption the derivation to follow goes through with no substantive change.

With canonical ensemble conditions of fixed $\{N_A,N_B,V,T\}$,  the canonical ensemble partition function $Z$ for the solution is

%=======================================================
% 2.3
\begin{equation}
\label{eq2.3}
Z(N_A,N_B,V,T) =  \frac{h^{-(3N_A+3N_B)} }{N_A!N_B!} \int d \boldsymbol{r}_{A}^{N_A} d \boldsymbol{p}_{A}^{N_A} d \boldsymbol{r}_{B}^{N_B} d \boldsymbol{p}_{B}^{N_B} \mbox{e}^{-\beta H(\boldsymbol{r}_{A}^{N_A},\boldsymbol{p}_{A}^{N_A},\boldsymbol{r}_B^{N_B},\boldsymbol{p}_B^{N_B})}~,
\end{equation}
where $h$ is Planck's constant, and here and throughout the paper the configurational integrals are over the volume $V$. Note that $Z$ is dimensionless. Because the momentum integrals in (\ref{eq2.3}) are Gaussian, the kinetic part of the partition function can be evaluated analytically giving

%========================================
%2.4
\begin{equation}
\label{eq2.4}
Z(N_A,N_B,V,T)  = \frac{\Lambda_{A}^{-3N_A} \Lambda_{B}^{-3N_B}}{N_A! N_B!} \int d\boldsymbol{r}_A^{N_A} d\boldsymbol{r}_B^{N_B} \mbox{e}^{- \beta U(\boldsymbol{r}_A^{N_A},\boldsymbol{r}_B^{N_B})}~,
\end{equation}
%=======================================
where $\Lambda_A$ is the thermal wavelengh of an A-molecule,

%==================================
%2.5
\begin{equation}
\label{eq2.5}
\Lambda_A = \frac{h}{(2 \pi m_A k_B T)^\frac{1}{2}}~.
\end{equation}
%===========================================
The Helmholtz free energy $F$ of the solution is calculated from the canonical partition function using

%==============================================
%2.6
\begin{equation} 
\label{eq2.6}
F(N_A,N_B,V,T) = -k_BT \ln Z(N_A,N_B,V,T)~. 
\end{equation}
%================================================

As described earlier, MM worked in the grand canonical ensemble \cite{mcmillan-mayer}. The grand canonical  partition function $\widetilde{ \widetilde{Z}}(\mu_A,\mu_B,V,T)$ is related to the canonical partition function $Z(N_A,N_B,V,T)$ by the double discrete Laplace transform

%=============================================
% 2.7
\begin{equation}
\label{eq2.7}
\widetilde{ \widetilde{Z}}(\mu_A,\mu_B,V,T) = \sum_{N_A} \sum_{N_B} \mbox{e}^{\beta \mu_A N_A} \mbox{e}^{\beta \mu_B N_B} Z(N_A,N_B,V,T)~,
\end{equation}
%==============================================
and the corresponding grand potential $\widetilde{ \widetilde{F}}(\mu_A,\mu_B,V,T) =  -k_BT \ln \widetilde{ \widetilde{Z}}(\mu_A,\mu_B,V,T)          $ is related to the Helmholtz free energy $F(N_A,N_B,V,T)$ by the double Legendre transform

%===============================================
% 2.8
\begin{equation}
\label{eq2.8}
\widetilde{ \widetilde{F}} = F - \mu_A N_A - \mu_B N_B~.
\end{equation}                     
%=======================================================
(In thermodynamic relations like (\ref{eq2.8}) we should strictly use the averages $\langle N_A \rangle$ and $\langle N_B \rangle$, but since the relative fluctuations in $N_A$ and $N_B$ are negligible for macroscopic systems no harm will come from using the simpler notation $N_A$ and $N_B$ for the average numbers.) 

Following Hill\cite{hill1960} we find it simpler to work in a semi-grand canonical ensemble (grand with respect to solvent only), with semi-grand canonical partition function $\widetilde{Z}(N_A,\mu_B,V,T)$ defined by the single Laplace transform of the canonical partition function $Z(N_A,N_B,V,T)$,

%=====================================================
% 2.9
\begin{equation}
\label{eq2.9}
\widetilde{Z}(N_A,\mu_B,V,T) = \sum_{N_B} \mbox{e}^{\beta \mu_B N_B} Z(N_A,N_B,V,T)~,
\end{equation} 
%====================================================
and corresponding semi-grand potential $\widetilde{F}(N_A,\mu_B,V,T) = -k_BT \ln \widetilde{Z}(N_A,\mu_B,V,T) $ related to the Helmholtz free energy $F(N_A,N_B,V,T)$ by the single Legendre transform
%==================================================
% 2.10
\begin{equation}
\label{eq2.10}
\widetilde{F} = F - \mu_B N_B~.
\end{equation} 
%===============================================
 As discussed in the Introduction, $\{N_A,\mu_B,T,V \}$ are the natural control variables for analysis of osmotic pressure experiments. From (\ref{eq2.10}) and the standard expression for the differential $dF$,
%====================================================
% 2.11
\begin{equation}
\label{eq2.11}
dF = -SdT - pdV + \mu_AdN_A + \mu_BdN_B ~, 
\end{equation}
%===================================================
where $S$ is the entropy, we find for $d \widetilde{F}$
%===================================
% 2.12
\begin{equation}
\label{eq2.12}
d \widetilde{F} = -SdT - pdV + \mu_A dN_A - N_Bd\mu_B~.
\end{equation}
%=================================================
From (\ref{eq2.12}) we obtain the pressure $p$ and solute chemical potential $\mu_A$ using
%
%======================================================
% 2.13
\begin{equation}
\label{eq2.13}
p =- \left(\frac{\partial  \widetilde{F}}{\partial V}\right)_{N_A,\mu_B,T}~~~~~~,~~~~~~~~\mu_A = \left(\frac{\partial \widetilde{F}}{\partial N_A}\right)_{\mu_B,V,T}~~.
\end{equation}
%==============================================================
%

We consider the ratio of solution/pure solvent semi-grand canonical partition funtions at the same $\{\mu_B,V,T\}$ state conditions, i.e. $\widetilde{Z}(N_A,\mu_B,V,T)/\widetilde{Z}_B(\mu_B,V,T)$, where the pure solvent semi-grand partition function $\widetilde{Z}_B(\mu_B,V,T) \equiv \widetilde{Z}(0,\mu_B,V,T)$ is  also a grand partition function and could just as well be denoted $\widetilde{\widetilde{Z}}_B(\mu_B,V,T) \equiv \tilde{\tilde{Z}}(-\infty,\mu_B,V,T)$. From (\ref{eq2.9}) and (\ref{eq2.4}) we can express these two partition functions as
%
%=============================================================
% 2.14
\begin{equation}
\label{eq2.14}
\widetilde{Z}(N_A,\mu_B,V,T) = \frac{\Lambda_A^{-3N_A} }{ N_A!} \int d \textbf{r}_A^{N_A} \mbox{e}^{-\beta U_{AA}} \sum_{N_B} \frac{z_B^{N_B}}{ N_B!} \int d\boldsymbol{r}_B^{N_B} \mbox{e}^{-\beta U_{BB}} \mbox{e}^{- \beta U_{AB}}~,
\end{equation}
%===========================================
%2.15
\begin{equation}
\label{eq2.15}
\widetilde{Z}_B(\mu_B,V,T) = \sum_{N_B} \frac{z_B^{N_B}}{ N_B!} \int d\boldsymbol{r}_B^{N_B} \mbox{e}^{-\beta U_{BB}}~,
\end{equation}
%==========================================
where $z_B = \mbox{e}^{\beta \mu_B}/\Lambda_B^3$. From the last two equations we see that the ratio $\widetilde{Z}(N_A,\mu_B,V,T)/\widetilde{Z}_B(\mu_B,V,T)$ can be written as
%
%=====================================
%2.16
\begin{equation}
\label{eq2.16}
\frac{\widetilde{Z}(N_A,\mu_B,V,T)}{\widetilde{Z}_B(\mu_B,V,T)} = \frac{ \Lambda_A^{-3N_A}   }{N_A!} \int d\boldsymbol{r}_A^{N_A} \mbox{e}^{-\beta U_{AA}}  \langle\mbox{e}^{-\beta U_{AB}}
\rangle_B~,
\end{equation} 
%==============================================
where the solvent grand-canonical average $\langle...
\rangle_B$ of a quantity which depends on $B$-coordinates but not $B$-momenta is
%
%===============================================
% 2.17
\begin{equation}
\label{eq2.17}
\langle...
\rangle_B~ = \frac{\sum_{N_B} \frac{z_B^{N_B}}{ N_B!} \int d\boldsymbol{r}_B^{N_B} \mbox{e}^{-\beta U_{BB}}(...)}{\sum_{N_B} \frac{z_B^{N_B}}{ N_B!} \int d\boldsymbol{r}_B^{N_B} \mbox{e}^{-\beta U_{BB}}}~. 
\end{equation}
%==================================================== 
In the averaging $\langle\mbox{e}^{-\beta U_{AB}}
\rangle_B$ in (\ref{eq2.16}) using (\ref{eq2.17}), the solute $A$-coordinates are held fixed, and the solvent $B$-coordinates are averaged over with normalized Boltzmann weight $\mbox{e}^{-\beta U_{BB}}$. In the final step we multiply and divide (\ref{eq2.16}) by $V^{N_A}$ and write it as
%e
%==============================================
% 2.18
\begin{equation} 
\label{eq2.18}
\frac{\widetilde{Z}}{\widetilde{Z}_B} = \frac{(V/\Lambda_A^3)^{N_A}}{ N_A!}\langle\mbox{e}^{-\beta U_
{AA}} \langle\mbox{e}^{-\beta U_{AB}}
\rangle_B
\rangle_{A,0}~,
\end{equation}
%====================================================
where $\langle...
\rangle_{A,0}$ denotes an unweighted average over the $A$-coordinates,
%
%=================================================
% 2.19
\begin{equation}
\label{eq2.19}
\langle...
\rangle_{A,0}~ = \int \frac{d \boldsymbol{r}_A^{N_A}}{V^{N_A}}(...)~.
\end{equation}
%===============================================

The solution-solvent difference of semi-grand potentials is $\widetilde{F} - \widetilde{F}_B = -k_BT \ln(\widetilde{Z}/\widetilde{Z}_B)$. From (\ref{eq2.18}), using the standard Stirling approximation for large numbers $\ln N_A! \doteq N_A \ln N_A - N_A $, we find for the dimensionless difference $\beta(\widetilde{F} - \widetilde{F}_B)$ the expression
%
%====================================================
% 2.20
\begin{equation}
\label{eq2.20}
\beta(\widetilde{F} - \widetilde{F}_B) = \beta F_A^{ideal} - \ln\langle\mbox{e}^{-\beta U_{AA}} \langle\mbox{e}^{-\beta U_{AB}}
\rangle_B
\rangle_{A,0}~,
\end{equation}
%=========================================================
where the ideal gas value of the solute Helmholtz free energy is given by
%
%===================================================
% 2.21
\begin{equation}
\label{eq2.21}
\beta F_A^{ideal} = N_A(\ln(c_A \Lambda_A^3)- 1)~,
\end{equation}
%===================================================
with $c_A = N_A/V$ the solute number density.
%
%
%
%
%
%
%
% % % % % % % % % % % % % % % % % % % % % % % % % % % % % % % % % % % % % % % % % % % % % % % % % % % % % % % % % % % % % % % % % % % % % % % % % % % % % % % % % % % % % % % % % % % %
% % % % % % % % % % % % % % % % % % % % % % % % % % % % % % % % % % % % % % % % % % % % % % % % % % % % % % % % % % % % % % % % % % % % % % % % % % % % % % % % % % % % % % % % % % % % %
%
\section{INTRODUCTION OF THE PMF's}
%
%
% % % % % % % % % % % % % % % % % % % % % % % % % % % % % % % % % % % % % % % % % % % % % % % % % % % % % % % % % % % % % % % % % % % % % % % % % % % % % % % % % % % % % % % % % % % % %
% % % % % % % % % % % % % % % % % % % % % % % % % % % % % % % % % % % % % % % % % % % % % % % % % % % % % % % % % % % % % % % % % % % % % % % % % % % % % % % % % % % % % % % % % % % % % %
%
The inner solvent average $\langle\mbox{e}^{-\beta U_{AB}}
\rangle_B $ in (\ref{eq2.20}) is a dimensionless positive function of the fixed $A$-configuration $\boldsymbol{r}_A^{N_A}$, which we write as
%
%==================================================================
% 3.1
\begin{equation}
\label{eq3.1}
\langle\mbox{e}^{-\beta U_{AB}}
\rangle_B~ \equiv~ \mbox{e}^{-\beta V_A}~,
\end{equation}
%=================================================================
which defines a solvent-induced effective $A$-potential $V_A(\boldsymbol{r}_A^{N_A})$. We form the total effective $A$-potential $W_A(\boldsymbol{r}_A^{N_A})$ by adding the vacuum solute potential $U_{AA}(\boldsymbol{r}_A^{N_A})$,
%
%============================================================
% 3.2
\begin{equation}
W_A(\boldsymbol{r}_A^{N_A}) = U_{AA}(\boldsymbol{r}_A^{N_A}) + V_A(\boldsymbol{r}_A^{N_A})~,
\end{equation}
%===========================================================
where  the total  vacuum $A$-potential $U_{AA}$ is defined in (\ref{eq2.2}) for our model. The complete average in (\ref{eq2.20}) then has the form
%
%========================================================
% 3.3
\begin{equation}
\label{eq3.3}
\langle\mbox{e}^{-\beta U_{AA}} \langle\mbox{e}^{-\beta U_{AB}}
\rangle_B
\rangle_{A,0} ~= \int \frac{d \boldsymbol{r}_A^{N_A}}{V^{N_A}} \mbox{e}^{-\beta W_A(\boldsymbol{r}_A^{N_A})}~.
\end{equation}
%=======================================================================
The right side of (\ref{eq3.3}) has the form of a solute-only canonical ensemble configurational  partition function, since the solvent $B$-coordinates have been integrated out  in defining the effective solute potential $W_A(\boldsymbol{r}_A^{N_A})$. This is the semi-grand canonical ensemble version of the MMI theorem discussed in the Introduction. In terms of the total effective solute potential $W_A$, using (\ref{eq3.3}) and (\ref{eq2.19}) we re-write the semi-grand potential (\ref{eq2.20}) in terms of the solute average $\langle\mbox{e}^{-\beta W_A}\rangle_{A,0}~$,

%==============================================
% % 4.3
\begin{equation}
\label{eq4.3}
\beta \tilde{F} = \beta \tilde{F}_B + \beta F_A^{ideal} - \ln \langle\mbox{e}^{-\beta W_A}
\rangle_{A,0}~.
\end{equation}

The total solute effective potential $W_A(\boldsymbol{r}_A^{N_A})$ is the PMF for the whole set of solute A molecules (see discussion later in this section). In our model, for simplicity  the total solute vacuum potential $U_{AA}$ has been assumed to be pairwise additive, $U_{AA}(\boldsymbol{r}_A^{N_A}) = \sum_{i<j} u_{AA}(r_{ij})$. In contrast, the total solute effective potential  $W_A(\boldsymbol{r}_A^{N_A})$ will necessarily contain one-body, two-body, three-body, etc, terms:
%
%========================================================
% 3.4
\begin{equation}
\label{eq3.4}
W_A(\boldsymbol{r}_A^{N_A}) = \sum_{i}w_A(\boldsymbol{r}_i) +\sum_{i < j}w_{AA}(r_{ij})+ \sum_{i<j<k} w_{AAA}(r_{ij},r_{ik},r_{jk}) +...~~.
\end{equation}
%============================================================
This is a cluster decomposition of the total effective potential $W_A(\textbf{r}_A^{N_A})$. We give next an intuitive derivation of (\ref{eq3.4}), with explicit expressions for the one- and two-body terms, and a few more details are given in Appendix A.

At very low solute concentrations (infinite dilution), we neglect pairs, triplets, etc, of $A$ molecules. In the solvent average $\langle\mbox{e}^{-\beta U_{AB}}
\rangle_B$ with fixed $A$-configuration, we assume the $A$-molecules are all isolated from each other and write the average as

%===================================================
%3.5
\begin{equation}
\label{eq3.5}
\langle\mbox{e}^{-\beta U_{AB}}
\rangle_B = \langle\mbox{e}^{-\beta U_{A_{1}B}} \mbox{e}^{-\beta U_{A_{2}B}} ...
\rangle_B \doteq \langle\mbox{e}^{-\beta U_{A_{1}B}}
\rangle_B \mbox{ }
\langle \mbox{e}^{-\beta U_{A_{2}B}}
\rangle_B... \equiv \mbox{e}^{-\beta v_{A_1}} \mbox{e}^{-\beta v_{A_2}} ...~,      
\end{equation}
%====================================================
where the contibutions $U_{A_{1}B},~ U_{A_{2}B}$,  etc, to $U_{AB} $ from molecules $A_1,A_2$, etc, in the average  (\ref{eq3.5}) have been assumed uncorrelated due to the assumption of far-apart $A$-molecules; in the pairwise additive vacuum potential model these contributions are given explicitly by  $U_{A_{1}B} \equiv \sum_{j} u_{AB}(r_{1j}),~ U_{A_{2}B} \equiv \sum_{j} u_{AB}(r_{2j})$, etc.                        The last step in (\ref{eq3.5}) simply defines the effective solute one-body potentials $v_{A_1} \equiv v_A(\boldsymbol{r}_1), v_{A_2} \equiv v_A(\boldsymbol{r}_2)$, etc, by

%====================================================
%3.6
\begin{equation}
\label{eq3.6}
\langle\mbox{e}^{-\beta U_{A_{1}B}}
\rangle_B~ \equiv \mbox{e}^{-\beta v_{A_1}}~.
\end{equation}
%===================================================
The average in (\ref{eq3.6}) involves the interaction with the solvent of just one solute molecule $A_1$ fixed at position $\boldsymbol{r}_1$, averaged over the solvent configurations. For $N_A$ identical solute molecules in a uniform solvent the effective one-body potentials $v_A(\boldsymbol{r}_i) $ are all equal and independent of position, so that the last expression in (\ref{eq3.5}) is the product of $N_A $ identical constant factors. The total one-body PMF term in (\ref{eq3.4}),~  $\sum_{i}w_A(\boldsymbol{r}_i) = \sum_{i}v_A(\boldsymbol{r}_i)$, is thus given by

%====================================================
% 3.7
\begin{equation}
\label{eq3.7}
\sum_{i}w_A(\boldsymbol{r}_i) = N_A v_{A_1}~,
\end{equation}
%===============================================
where $v_{A_1}$ is the constant one-body  potential defined by (\ref{eq3.6}). Equation (\ref{eq3.6}) defining $v_{A_1}$ was first derived by Widom \cite{widom1963} with potential distribution theory using the canonical ensemble, and its physical significance is discussed further later.

We now consider a slightly increased solute concentration so that solute-solute pair interactions, but no triplets etc, also need to be considered. Suppose the fixed $A$-configuration has the $A_1A_2$ pair and a few other solute pairs close together, and the other solute molecules  isolated. Just as (\ref{eq3.5}) factored into a product of uncorrelated one-body terms at infinite dilution, now the average $\langle\exp{-\beta U_{AB}}
\rangle_B$ will factor into a product of uncorrelated two-body terms such as $\langle\exp-\beta(U_{A_{1}B} + U_{A_{2}B})
\rangle_B$ and other such terms arising from the close $A$-pairs, and constant factors coming from the uncorrelated and isolated $A$-molecules. We denote the $A_1A_2$ pair term by

%=======================================================
%3.8
\begin{equation}
\label{eq3.8}
\langle\mbox{e}^{-\beta(U_{A_{1}B} + U_{A_{2}B} )}
\rangle_B~ \equiv \mbox{e}^{-\beta (v_{A_{1}} + v_{A_{2}} + v_{A_{1}A_{2}}) }~.
\end{equation}
%=====================================================
According to (\ref{eq3.8}) we now fix two solute molecules $A_1$ and $A_2$ at positions $\boldsymbol{r}_1$ and $\boldsymbol{r}_2$ in the otherwise pure solvent, and average the exponential of $ U_{A_{1}B} + U_{A_{2}B}$ over solvent configurations. If $A_1$ and $A_2$ are far apart as assumed previously, the result is $\exp-\beta(v_{A_1} + v_{A_2})$ as before. When $A_1$ and $A_2$ are close together, the result is modified and denoted by the right side of (\ref{eq3.8}), thereby defining a solvent-induced contribution $v_{A_{1}A_{2}}$ to the total effective $A_{1}A_{2}$ pair potential $w_{A_{1}A_{2}} = u_{A_{1}A_{2}} + v_{A_{1}A_{2}}$, where $u_{A_{1}A_{2}} \equiv u_{AA}(r_{12}) $ is the vacuum pair potential. Thus $v_{A_{1}A_{2}} \equiv v_{AA}(r_{12})$ is the extra contribution, beyond $v_A(\boldsymbol{r}_1) + v_A(\boldsymbol{r}_2)$, to the solvent-induced effective potential of solute molecules $A_1$ and $A_2$, and  is therefore normalized such that $v_{AA}(r_{12}) \rightarrow 0 $ for $r_{12} \rightarrow \infty$, the same normalization as the vacuum pair potential $u_{AA}(r_{12})$. The total effective $A_{1}A_2 $ pair potential $w_{AA}(r_{12}) = u_{AA}(r_{12}) + v_{AA}(r_{12})$ thus has the same normalization. The total effective pairwise solute potential is a sum of terms like $w_{AA}(r_{12})$ as written in (\ref{eq3.4}). Equation (\ref{eq3.8}) is easily related to one derived in potential disribution theory \cite{widom1963}  for the solute-solute pair correlation function at infinite dilution $g_{AA}(r_{12})$, using (see below) $g_{AA}(r_{12})= \exp-\beta w_{AA}(r_{12})$. This brief derivation of the effective pairwise term is elaborated in Appendix A. 

In the next section we will  consider virial expansions to $O(c_A^2)$ so that we will not need to consider the triplet $w_{AAA}(r_{12},r_{13},r_{23})$, etc, effective solute potentials in (\ref{eq3.4}) in this paper, but it is clear how they can be generated by continuing the pattern $w_{A}(\boldsymbol{r_1}), w_{AA}(r_{12})$, ...~.

The total effective solute potential $ W_{A}(\boldsymbol{r}_A^{N_A})= V_A(\boldsymbol{r}_A^{N_A})+U_{AA}(\boldsymbol{r}_A^{N_A})$, and its component $n$-body effective potentials        $w_A(\boldsymbol{r}_1)=v_A(\boldsymbol{r}_1)$, $ w_{AA}(r_{12}) = v_{AA}(r_{12}) + u_{AA}(r_{12}),~  w_{AAA}(r_{12},r_{13},r_{23}) = v_{AAA}(r_{12},r_{13},r_{23}),~ ...~ $, are solute PMFs. These are reversible works, or changes in constrained free energy (here the semi-grand potential), when, starting from rest and far apart in the vacuum, we add solute molecules at fixed positions to the pure solvent at a given chemical potential and temperature. Thus $w_A(\boldsymbol{r}_1)$ is the reversible work to add a solute molecule at position $\boldsymbol{r}_1$ in the pure solvent, with fixed state conditions $\{\mu_B, T\}$. For this case, where we add just one solute molecule to the pure solvent, the work is a solvent intensive property and we can just as well regard $\{p,T\}$ or $\{c_B,T\}$ as fixed, so that this reversible work is also equal to the changes in Gibbs and Helmholtz free energies.  Like all PMFs \cite{hill1960}, the gradient gives the  mean force on a molecule; here $-\boldsymbol{\nabla}_1w_A(\boldsymbol{r}_1)$ gives the solvent-averaged mean force  on a solute molecule fixed at $\boldsymbol{r}_1$. In this case the mean force vanishes, as it should for a uniform solvent, as $w_A(\boldsymbol{r}_1)$ is independent of $\boldsymbol{r}_1$. Similarly, the total reversible work to add a pair of solute molecules $A_1$ and $A_2$ to the pure solvent at positions $\boldsymbol{r}_1$ and $\boldsymbol{r}_2$, starting with $A_1$ and $A_2$ at rest and  far apart in the vacuum, is $w_A(\boldsymbol{r}_1)+ w_A(\boldsymbol{r}_2)+ w_{AA}(r_{12})$. The terms $w_A(\boldsymbol{r}_1)+ w_A(\boldsymbol{r}_2)$ give the reversible work to add $A_1$ and $A_2$ to the solvent keeping them far apart, and $ w_{AA}(r_{12})$ is the reversible work to bring them to finite separation $r_{12}$ in the solvent, starting far apart. For the pair PMF $w_{AA}(r_{12})$,~ $-\boldsymbol{\nabla}_1w_{AA}(r_{12})$ gives the solvent-averaged mean force on solute molecule $A_1$ fixed at $\boldsymbol{r}_1$, with $A_2$ fixed at $\boldsymbol{r}_2$. This mean force will be nonvanishing if the separation $r_{12}$ between $A_1$ and $A_2$  is not too large. Part of the mean force is the vacuum force $-\boldsymbol{\nabla}_1u_{AA}(r_{12}) $ and the other part $-\boldsymbol{\nabla}_1v_{AA}(r_{12})$ is induced by the solvent. PMFs can be expressed in terms of the logarithm of the corresponding correlation functions\cite{hill1960}. Thus for $w_{AA}(r_{12})$ we have $ g_{AA}(r_{12}) = \exp-\beta w_{AA}(r_{12})$, where $g_{AA}(r_{12})$ is the solute-solute pair correlation function for an infinitely dilute solution.

For notational simplicity we have assumed monatomic solute and solvent molecules. For polyatomic molecules, which may be nonrigid, besides the obvious change\cite{gray-gubbins} required for  the solute ideal gas free energy (\ref{eq2.21}), one need only include the internal coordinates of the solute and solvent molecules in the averaging over the $B$-coordinates when forming the total effective solute potential for the solute molecular centers $W_A(\boldsymbol{r}_A^{N_A})$, and its one-body and two-body components  $w_A(\boldsymbol{r}_1)$ and $w_{AA}(r_{12})$  calculated using (\ref{eq3.6}) and (\ref{eq3.8}). The internal degrees of freedom effectively all become part of the solvent to be integrated out, in forming the effective solute potentials which depend only on the coordinates of the fixed  solute molecular centers (usually chosen to be the centers of mass).

% % % % % % % % % % % % % % % % % % % % % % % % % % % % % % % % % % % % % % % % % % % % % % 
% % % % % % % % % % % % % % % % % % % % % % % % % % % % % % % % % % % % % % % % % % % % % % 
% % % % % % % % % % % % % % % % % % % % % % % % % % % % % % % % % % % % % % % % % % % % % % %
% % % % % % % % % % % % % % % % % % % % % % % % % % % % % % % % % % % % % % % % % % % % % % %
%
\section{VIRIAL EXPANSIONS}
%
% % % % % % % % % % % % % % % % % % % % % % % % % % % % % % % % % % % % % % % % % % % % % % % %
% % % % % % % % % % % % % % % % % % % % % % % % % % % % % % % % % % % % % % % % % % % % % % % % % % %
% % % % % % % % % % % % % % % % % % % % % % % % % % % % % % % % % % % % % % % % % % % % % % % % % %
% % % % % % % % % % % % % % % % % % % % % % % % % % % % % % % % % % % % % % % % % % % % % % % % % %

Here we assume a dilute solution and  derive the virial expansions for various properties. We first employ a simple canonical ensemble technique, developed for gas virial expansions by van Kampen\cite{vankampen} (see also Landau and Lifshitz\cite{landau-lifshitz}), to derive the virial series in solute concentration $c_A =N_A/V$ for the change of semi-grand potential upon adding $N_A$ solute molecules to  pure solvent at fixed solvent chemical potential, volume, and temperature, $\tilde{F}(N_A,\mu_B,V,T) -\tilde {F}_B(\mu_B,V,T)$. Because our effective solute partition function is canonical, we can use the van Kampen  canonical ensemble method and thereby  avoid the complications of the standard grand canonical ensemble method; the concentration expansion is derived directly with  no need for the introduction and subsequent elimination of an activity expansion, and no graphs at all are needed, which avoids the problem of defining the graphs, with their various types  such as reducible and irreducible, and then showing that the net result is due only to the irreducible ones. 
From  this basic virial series, we then  use thermodynamic arguments  to derive virial series for the solute and solvent chemical potentials, $\mu_A(c_A,\mu_B,T)$, $\mu_A(c_A,p,T)$ and $\mu_B(c_A,p,T)$, the osmotic pressure $\pi = p(c_A,\mu_B,T) - p_B(\mu_B,T)$, and  the changes of the Gibbs free energy and enthalpy, $G(N_A,N_B,p,T) - G_B(N_B,p,T)$ and $H(N_A,N_B,p,T) - H_B(N_B,p,T)$, respectively. As is clear from the notation, some of these derived series involve changing independent variables as well as the dependent one.

% % % % % % % % % % % % % % % % % % % % % % % % % % % % % % % % % % % % % % % % % % % % % % % % % % %
% % % % % % % % % % % % % % % % % % % % %
%
\subsection{Virial expansion for semi-grand potential difference $\tilde{F} - \tilde{F}_B$}
%
% % % % % % % % % % % % % % % % % % % % %
% % % % % % % % % % % % % % % % % % % % % % % % % % % % % % % % % % % % % % % % % % % % % % % % % % % % %

The semi-grand potential is an extensive quantity, and  we now assume low solute concentration and derive a virial series for $\tilde{F} -\tilde{F}_B - F_A^{ideal} $ to $ O(N_A^2/V)$, i.e.,

% % % % % % % % % % % % % % % % % % % % % % % % % % % % % % % % % % % % % % % % %
% 4.1
\begin{equation}
\label{eq4.1}
\beta\tilde{F}(N_A,\mu_B,V,T) = \beta \tilde{F}_B(\mu_B,V,T)+ \beta F_A^{ideal} + \tilde{A}(\mu_B,T)N_A + \tilde{B}(\mu_B,T)\frac{N_A^2}{V}~,
\end{equation}
% % % % % % % % % % % % % % % % % % % % % % % % % % % % % % % % % %
where $F_A^{ideal}$ is defined in (\ref{eq2.21}), and $\tilde{A}$ and $\tilde{B}$ are the first and second virial coefficients for the quantity $\beta \tilde{F} - \beta \tilde{F}_B - \beta F_A^{ideal}$. Note that $\tilde{A}$ is dimensionless and $\tilde{B}$ has dimensions of volume. The virial coefficients to be defined later for other properties will also have these dimensions for the most part.

From the virial series (\ref{eq4.1}) for $\tilde{F}$  we immediately get one for the solute chemical potential $\mu_A$ by differentiating with respect to $N_A$ (see the second thermodynamic relation in (\ref{eq2.13})),

% % % % % % % % % % % % % % % % % % % % % % % % % % %
% 4.2
\begin{equation}
\label{eq4.2}
\beta \mu_A(c_A,\mu_B,T) = \beta \mu_A^{ideal} + \tilde{A}(\mu_B,T) + 2 \tilde{B}(\mu_B,T)c_A~,
\end{equation}
% % % % % % % % % % % % % % % % % % % % % % % % % % % %
where $ \beta \mu_A^{ideal} = \ln(c_A \Lambda_A^3)$ is the ideal monatomic gas value (which will be different for polyatomic solutes). The solute chemical potential is an intensive property and for a binary solution can be expressed in terms of three independent intensive variables, here chosen to be $\{c_A,\mu_B,T\}$.

The virial series (\ref{eq4.1}) will now be derived using (\ref{eq4.3}) from an analysis of the low order solute clusters which dominate the average $\langle\mbox{e}^{-\beta W_A}\rangle_{A,0}$ at low solute concentrations.  At very low solute concentration (infinite dilution), essentially only isolated solute molecules occur and thus only the one-body term $W_A^{(1)} \equiv \sum_{i}w_A(\boldsymbol{r}_i)$ in $W_A$ will matter, as pair, etc, interactions are negligible. Because the one-body term is independent of the $A$-coordinates (see (\ref{eq3.7})) the averaging in (\ref{eq4.3}) is redundant and a contribution to $\tilde{F}$ linear in $N_A$ results, with first virial cofficient $\tilde{A}$ given by

% % % % % % % % % % % % % % % % % % % % % % % % % % %
% 4.3
%\begin{equation}
%\label{eq4.3}
%\beta \tilde{F} = \beta \tilde{F}_B + \beta F_A^{ideal} - \ln \langle\mbox{e}^{-\beta W_A}
%\rangle_{A,0}~,
%\end{equation}
% % % % % % % % % % % % % % % % % % % % % % % % % % % % % % % % %
%where the total effective solute potential $W_A$ is given by (\ref{eq3.4})

%At very low solute concentration (infinite dilution), only the one-body term $W_A^{(1)} \equiv %\sum_{i}w_A(\boldsymbol{r}_i)$ in $W_A$ will matter, as pair etc interactions are negligible. Because %the one-body term is independent of the $A$-coordinates (see (\ref{eq3.7})) the averaging in %(\ref{eq4.3}) is trivial and a contribution to $\tilde{F}$ linear in $N_A$ results, with first virial %cofficient $\tilde{A}$ given by

% % % % % % % % % % % % % % % % % % % % % % % % % % % %
% 4.4
\begin{equation} 
\label{eq4.4}
\tilde{A}(\mu_B,T) = \beta v_{A_1}~,
\end{equation}
% % % % % % % % % % % % % % % % % % % % % % % % % % % % % % %
where $v_{A_1}$ is defined by (\ref{eq3.6}). The physical significance of $v_{A_1}$ is now apparent from (\ref{eq4.2}) and (\ref{eq4.4}): $v_{A_1}$ defined by (\ref{eq3.6}) is the solute excess chemical potential at infinite dilution, $\mu_A - \mu_A^{ideal}$ for $c_A \rightarrow 0 $, as first shown by Widom\cite{widom1963} with potential distribution theory  using the canonical ensemble.

We now consider slightly increased solute concentrations such that essentially only solute singlets and pairs occur, and thus one-body and two-body, but not three-body, etc, solute interaction terms in $W_A(\boldsymbol{r}_A^{N_A})$  need to be considered, i.e., the first two terms in (\ref{eq3.4}). For these one-body and two-body terms  the total effective potential is  $W_A = W_A^{(1)} + W_{AA}$, where  $W_A^{(1)} \equiv \sum_{i} w_{A_i}$ and $W_{AA} \equiv \sum_{i<j} w_{A_iA_j} $, and the average in (\ref{eq4.3}) can  be written

%===========================================================
% 4.5
\begin{equation}
\label{eq4.5}
 \langle\mbox{e}^{-\beta W_A}
\rangle_{A,0}~ = \mbox{e}^{-\beta W_A^{(1)}}\langle\mbox{e}^{-\beta W_{AA}}
\rangle_{A,0}~,
\end{equation}
%===========================================================
since $W_A^{(1)}$ is independent of the $A$-coordinates. When we take the logarithm of the expression (\ref{eq4.5}), $W_A^{(1)}$ will contribute the first virial term to (\ref{eq4.3}) as before, and we focus now on the pairs average ~~$\langle\mbox{e}^{-\beta W_{AA}}
\rangle_{A,0}$ in (\ref{eq4.5}), which will be seen to generate the second virial term. Written out in full, this average is

%================================================
% 4.6
\begin{equation}
\label{eq4.6}
\langle\mbox{e}^{-\beta W_{AA}}
\rangle_{A,0}~ \equiv \langle\mbox{e}^{-\beta w_{A_1A_2}}~ \mbox{e}^{-\beta w_{A_1A_3}}~\mbox{e}^{-\beta w_{A_2A_3}}...
\rangle_{A,0}~.
\end{equation}\
%===============================================
Because of the neglect of triplet, etc, solute configurations, the pair terms $w_{A_1A_2}, w_{A_1A_3}$, etc., are uncorrelated in the average (\ref{eq4.6}), so that we have approximately

%==============================================
% 4.7
\begin{equation}
\label{eq4.7}
\langle\mbox{e}^{-\beta W_{AA}}
\rangle_{A,0}~\doteq \langle\mbox{e}^{-\beta w_{A_1A_2}}
\rangle_{A,0}~\langle\mbox{e}^{-\beta w_{A_1A_3}}
\rangle_{A,0}~\langle\mbox{e}^{-\beta w_{A_2A_3}}
\rangle_{A,0}...~.
\end{equation}
%=================================================
Since all $N_A(N_A - 1)/2$  pairs are equivalent on average by symmetry, we have

%========================================= 
% 4.8
\begin{equation}
\label{eq4.8}
\langle\mbox{e}^{-\beta W_{AA}}
\rangle_{A,0}~\doteq  [\langle\mbox{e}^{-\beta w_{A_1A_2}}
\rangle_{A,0} ]^{N_A(N_A-1)/2}~.
\end{equation}
%================================================
For a uniform fluid the averaging on the right side can be taken over the relative position $\boldsymbol{r}_{12}$ of molecule $A_2$ with respect to $A_1$, and because the fluid is isotropic the effective pair potential $w_{A_1A_2}$ depends only on the magnitude $r_{12}$ of $\boldsymbol{r}_{12}$.  We assume $ w_{A_1A_2}$ is short ranged so that the quantity on the right side being averaged,  $\mbox{e}^{-\beta w_{A_1A_2}} $,  differs from unity only in a small part of the complete averaging region $V$, and we therefore     introduce the two-body Mayer cluster function $\mbox{e}^{-\beta w_{A_1A_2}} - 1$ which differs from zero only in the small region,  and write (\ref{eq4.8}) as

%================================================
% 4.9
\begin{equation}
\label{eq4.9}
\langle\mbox{e}^{-\beta W_{AA}}
\rangle_{A,0}~\doteq [1~ + \langle\mbox{e}^{-\beta w_{A_1A_2}} - 1
\rangle_{A,0}]^{N_A(N_A-1)/2}~.
\end{equation}
%===============================================
We need $ \ln \langle\mbox{e}^{-\beta W_{AA}}
\rangle_{A,0}$  and by using $ \ln(1+x)^n = n \ln(1+x),~ \ln(1+x) \doteq x $  for $x\ll 1$ (here $x$ is $O(1/V)$), and $n \equiv  N_A(N_A-1)/2 \doteq N_A^2/2$ for $N_A\gg 1$, we have 

%===========================================
% 4.10
\begin{eqnarray}
\label{eq4.10}
\ln \langle\mbox{e}^{-\beta W_{AA}}
\rangle_{A,0}~ &\doteq& \frac{1}{2}N_A^2 \langle\mbox{e}^{-\beta w_{A_1A_2}}-1
\rangle_{A,0} \nonumber\\
&\equiv& -\frac{N_A^2}{V} \tilde{B}(\mu_B,T)~,
\end{eqnarray}
%============================================
where

%=============================================
% 4.11
\begin{equation}
\label{eq4.11}
\tilde{B}(\mu_B,T) = -\frac{1}{2} \int d\boldsymbol{r}_{12} (\mbox{e}^{-\beta w_{AA}(r_{12})}-1)~.
\end{equation}
%===============================================
Using (\ref{eq4.10}) and (\ref{eq4.5}) in (\ref{eq4.3}) now gives the second virial term in (\ref{eq4.1}), with osmotic second virial coefficient given by (\ref{eq4.11}).

We will not be needing them, but molecular expressions for the third and higher osmotic virial coefficients $\tilde{C}(\mu_B,T)$, etc, can also be derived by this simple method, parallelling the results derived for the higher gas virial coefficients by this method \cite{vankampen,oppenheim-terwiel,ramshaw}.

% % % % % % % % % % % % % % % % % % % % % % % % % % % % % % % % % % % % % % % % % % % % % % % % % %
% % % % % % % % % % % % % % % % % % %
% 
\subsection{Virial expansion for osmotic pressure $\pi$}
% 
% % % % % % % % % % % % % % % % % % % 
% % % % % % % % % % % % % % % % % % % % % % % % % % % % % % % % % % % % % % % % % % % % % % % % %
We find the virial series for the osmotic pressure $\pi = p(c_A,\mu_B,T) - p_B(\mu_B,T)$ by differentiating the series (\ref{eq4.1}) for the semi-grand potential $\tilde{F}$  with respect to volume (see the first thermodynamic relation in (\ref{eq2.13})). For $p$ we thus have  

%======================================================= 
% 4.12
\begin{equation}
\label{eq4.12}
\beta p(c_A,\mu_B,T) = \beta p_B(\mu_B,T)  + c_A + \tilde{B}(\mu_B,T)c_A^2~,
\end{equation}
%===========================================================
from which we immediately get the series (\ref{eq1.1}) for $\pi$.

Equation (\ref{eq4.12}) can also be derived from the virial  series (\ref{eq4.2}) for $\mu_A$ and the Gibbs-Duhem relation which we will need later, 

%======================================
% 4.13
\begin{equation}
\label{eq4.13}
dp = c_A d\mu_A + c_B d\mu_B~,~~~~~~~~~~~~~(T~ fixed)~,
\end{equation}
%===========================================
from which we get at fixed $\mu_B$ and $T$,

%====================================================
%4.14
\begin{equation}
\label{eq4.14}
\left(\frac{\partial p}{\partial c_A}\right)_{\mu_B,T} = c_A\left(\frac{\partial \mu_A}{\partial c_A}\right)_{\mu_B,T}~.
\end{equation}
%=====================================================
It is easy to check that the series (\ref{eq4.12})  and (\ref{eq4.2}) are consistent with (\ref{eq4.14}).

% % % % % % % % % % % % % % % % % % % % % % % % % % % % % % % % % % % % % % % % % % % % %
% % % % % % % % % % % % % % % % % % % % % % % % % % % % % % % % % %
%
\subsection{Virial expansions for other thermodynamic properties}
%
% % % % % % % % % % % % % % % % % % % % % % % % % % % % % % % % % %
% % % % % % % % % % % % % % % % % % % % % % % % % % % % % % % % % % % % % % % % % % % % % %

There has been some earlier work on extending MM theory to obtain virial expansions for other solution properties. Hill \cite{hill1957} uses the Stockmayer semi-grand ensemble with fixed $ \{\mu_A,N_B,p,T \}$
to obtain virial series for the chemical potentials, Gibbs free energy and enthalpy. Friedman \cite{friedman} derives the relation between intensive properties expressed as a function of MM variables $\{c_A,p_B,T\}$ and the more experimentally accessible variables $\{c_A,p,T\}$. Kozak et al \cite{kozak-et-al} discuss the solvent chemical potential $\mu_B$ and the enthalpy of mixing within the MM framework,  and Rossky and Friedman \cite{rossky-friedman} derive  the second virial correction to the Henry law ``constant'' $K_H$, which is a measure of (in)solubility of the $A$ species. 

Using our semi-grand canonical ensemble with fixed $\{N_A,\mu_B,V,T\}$ we have obtained above the virial series for semi-grand potential $\tilde{F}$, solute chemical potential $\mu_A$, and pressure $p$, and we now derive virial series for other properties from these using purely thermodynamic arguments. We will also change independent variables in some cases to match the most convenient experimental variables. For example, we will discuss the enthalpy $H$ as a function of the independent or control variables $\{N_A,N_B,p,T \}$ used in ITC experiments.

The enthalpy $H$ as a function of $\{N_A,N_B,p,T\}$ can be obtained from the Gibbs free energy $G(N_A,N_B,p,T)$ using the Gibbs-Helmholtz relation

%===========================================================
% 4.15
\begin{equation}
\label{eq4.15}
\beta H = -T \left(\frac {\partial \beta G}{\partial T} \right)_{N_A,N_B,p}~,
\end{equation}
%===================================================
where we find $G$ from

%=======================================================
% 4.16
\begin{equation}
\label{eq4.16}
G = \mu_AN_A + \mu_BN_B~.
\end{equation}
%===================================================
The extensivity of $G$ is manifested  in (\ref{eq4.16}) by the factors $N_A$ and $N_B$ and we focus on the intensive chemical potentials $\mu_A$ and $\mu_B$, which are functions of three independent intensive variables. In (\ref{eq4.2}) we have a virial series in $c_A$ for  $\mu_A$ at constant $\mu_B$ and $T$. We first use thermodynamics to convert this $\mu_A$ series to one at constant $p$ and $T$, and from this new $\mu_A$ series  we will use thermodynamics to obtain a $\mu_B$ series at constant $p$ and $T$. 

Figure 1 shows what is to be done to change control variable from $\mu_B$ to $p$ in the $\mu_A$ series. The series (\ref{eq4.2}) for $\mu_A(c_A,\mu_B,T)$ is an expansion   along the curve $\mu_B = const $ in Fig.1, starting from the initial point (pure solvent) with $c_A = 0$. We now transform this expansion to one along the curve $p = const$, and, to $O(c_A)$, it will take the form

%========================================================
% % 4.17
\begin{equation}
\label{eq4.17}
\beta \mu_A(c_A,p,T) = \beta \mu_A^{ideal} + \hat{A}(p,T) +  \hat{B}(p,T)c_A~.
\end{equation}
%====================================================
The new virial coefficients $\hat{A}$ and $\hat{B}$ are Taylor series expansion coefficients  defined as usual by

%=============================================================
% % 4.18
\begin{equation}
\label{eq4.18}
\hat{A} = \beta \mu_A^{ex}(c_A,p,T)\biggr \rvert_{c_A\rightarrow0}~~,~~\hat{B}=\frac{\partial \beta \mu_A^{ex}}{\partial c_A}\biggr \rvert_{p,T,c_A \rightarrow0}~,
\end{equation}
%======================================================
where $\mu_A^{ex} = \mu_A - \mu_A^{ideal}$ is the solute excess chemical potential. We will relate these new coefficients to the old ones defined by Taylor series (\ref{eq4.2}),

%======================================================
% % 4.19
\begin{equation}
\label{eq4.19}
\tilde{A} = \beta \mu_A^{ex}(c_A,\mu_B,T)\biggr \rvert_{c_A\rightarrow0}~~,~~2\tilde{B}=\frac{\partial \beta \mu_A^{ex}}{\partial c_A}\biggr \rvert_{\mu_B,T,c_A \rightarrow0}~.
\end{equation}
%========================================================

The quantities $\hat{A}$ and $\tilde{A}$ are equal since they both are equal to the solute excess chemical potential in the infinite dilution limit $c_A\rightarrow0$; in this limit $\mu_A^{ex}$  depends on just the pure solvent independent intensive conditions and these can be chosen as $\{p,T\}$ or $\{\mu_B,T\}$, or any other suitable pair such as $\{c_B,T\}$. Thus our first relation is simply

%=============================================
% % 4.20
\begin{equation}
\label{eq4.20}
\hat{A} = \tilde{A}~.
\end{equation}
%==============================================

To derive the relation between $\hat{B}$ and $\tilde{B}$ we use the rule for partial derivatives to re-write the definition in (\ref{eq4.18}) in terms of a  derivative at constant $\mu_B$  in place of one at constant $p$:

%================================================
% % 4.21
\begin{equation}
\label{eq4.21}
\hat{B} = \left(\frac{\partial \beta \mu_A^{ex}}{\partial c_A} \right)_{\mu_B} + \left(\frac{\partial \beta \mu_A^{ex}}{\partial \mu_B} \right)_{c_A} \left(\frac{\partial \mu_B}{\partial c_A}\right)_p~,
\end{equation}
%=====================================================
where it is to be understood that $T$ is constant and the limit $c_A \rightarrow0$ is to be taken in each partial derivative here. The first partial derivative on the right side is $2\tilde{B}$ as seen from (\ref{eq4.19}). The second partial derivative in (\ref{eq4.21}) can be written as $\beta(\partial \mu_A/ \partial \mu_B)_{c_A}$ since $\mu_A^{ideal}$ does not depend on $\mu_B$, and $(\partial \mu_A/\partial \mu_B)_{c_A}$  is evaluated  in Appendix B (see (\ref{eqC.1d})), giving

%======================================
% % 4.22
\begin{equation}
\label{eq4.22}
\left(\frac{\partial \mu_A}{\partial \mu_B}\right)_{c_A} = c_B^0(\overline{v}_A^{\infty} - k_BT \chi_B) ~~~,~~~(c_A\rightarrow0)~,
\end{equation}
%==========================================
where $ \overline{v}_A^{\infty} $ is the solute partial molecular volume in the infinite dilution limit, $c_B^0$ is the pure solvent density, and $\chi_B$ is the pure solvent isothermal compressibility.  The third partial derivative  in (\ref{eq4.21}) can be transformed to the following form with the help of the Gibbs-Duhem relation (\ref{eq4.13}) with fixed $p$ and $T$:

%===============================================
% % 4.23
\begin{equation}
\label{eq4.23}
\left(\frac{\partial \mu_B}{\partial c_A}\right)_p = - \frac{c_A}{c_B} \left(\frac{\partial \mu_A}{\partial c_A}\right)_p~,~~~(p,T~ fixed)~.                    
\end{equation}
%=================================================
In the limit $c_A \rightarrow0$, $(\partial \mu_A/\partial c_A)_p$ on the right side is dominated by the ideal term $\mu_A^{ideal} = k_BT \ln(c_A \Lambda_A^3)$ of the virial series (\ref{eq4.17}), so that  $(\partial \mu_A/\partial c_A)_p \doteq k_BT/c_A $, and hence

%=====================================================
% % 4.24
\begin{equation}
\label{eq4.24}
\left(\frac{\partial \mu_B}{\partial c_A}\right)_p = -\frac{k_BT}{c_B^0}~,~~~(c_A\rightarrow0)~.
\end{equation} 
%===================================================
Equation (\ref{eq4.24}) is a version of Raoult's law, and can also be derived from the usual formulation of that law. Gregorio and Widom \cite{gregorio-widom} derive  (\ref{eq4.24})  from potential distribution theory.  Putting it all together gives the relation we seek:

%=========================================
% % 4.25
\begin{equation}
\label{eq4.25}
\hat{B}= 2\tilde{B} - \overline{v}_A^{\infty} + k_BT \chi_B ~. 
\end{equation}
%===============================================

Using (\ref{eq4.25}) and (\ref{eq4.20}) we write the first two terms of the constant pressure virial series (\ref{eq4.17}) for $\mu_A$ in terms of the original virial coefficients as 

%================================================
% % 4.26
\begin{equation}
\label{eq4.26}
\beta \mu_A(c_A,p,T)= \beta \mu_A^{ideal} + \tilde{A}(p,T) + \left(2 \tilde{B}(p,T) - \overline{v}_A^{\infty}(p,T)+k_BT \chi_B(p,T) \right)c_A~,
\end{equation}
%===========================================
where the original virial coefficients $\tilde{A}$ and $\tilde{B}$, which are  intensive solvent properties  defined in equations (\ref{eq4.4}) and (\ref{eq4.11}) and expressed there as  functions of chemical potential $\mu_B$,  are now to be re-expressed as  functions of pressure $p$. We see that the first virial coefficient for  $\mu_A$ in the series at constant pressure is the same as the first virial coefficient $\tilde{A}$  in the $\mu_A$  series at constant solvent chemical potential, but the new second virial coefficient  differs from $2 \tilde{B}$ which occurs in the original series. 
In general we expect the magnitudes of $\tilde{B}$ and $ \overline{v}_A^{\infty} $ to be of the same order and $k_BT \chi_B$ to be negligible (e.g., it is of order $1 \AA^3 $ for water at normal state conditions) and thus, for example, in cases where  $\tilde{B} < 0$ and  $ \overline{v}_A^{\infty} > 0$, the second virial coefficient of $\mu_A$ at constant pressure, $ 2\tilde{B} - \overline{v}_A^{\infty} +k_BT \chi_B$, will be negative and larger in magnitude than the second virial coefficient of $\mu_A$ at constant solvent chemical potential, $2\tilde{B}$. Exceptional situations can arise near a solvent theta temperature, where $\tilde{B}= 0$, and near a solvent liquid-gas critical point, where $\tilde{B}, ~ \overline{v}_A^{\infty}$, and $\chi_B$ can all become large.

  %Along the curve $p = const$ $\mu_B$ will change. We need the change thereby induced in the %coefficient $\tilde{A}(\mu_B,T)$ to $O(c_A)$, which will contribute to a new second virial %coefficient %for the constant-$p$ series. We will not need to consider the change induced in the %coefficient %$\tilde{B}(\mu_B,T)$ as such a change would generate an $O(c_A^2)$ term in $\mu_A$, %which we do not %consider. Thus, to first-order in $c_A$, the required change $\Delta \tilde{A}$ is

%	=========================================================================

\begin{figure}
\centering
%\DeclareGraphicsExtensions{.png, .bb}
\includegraphics[scale=0.5]{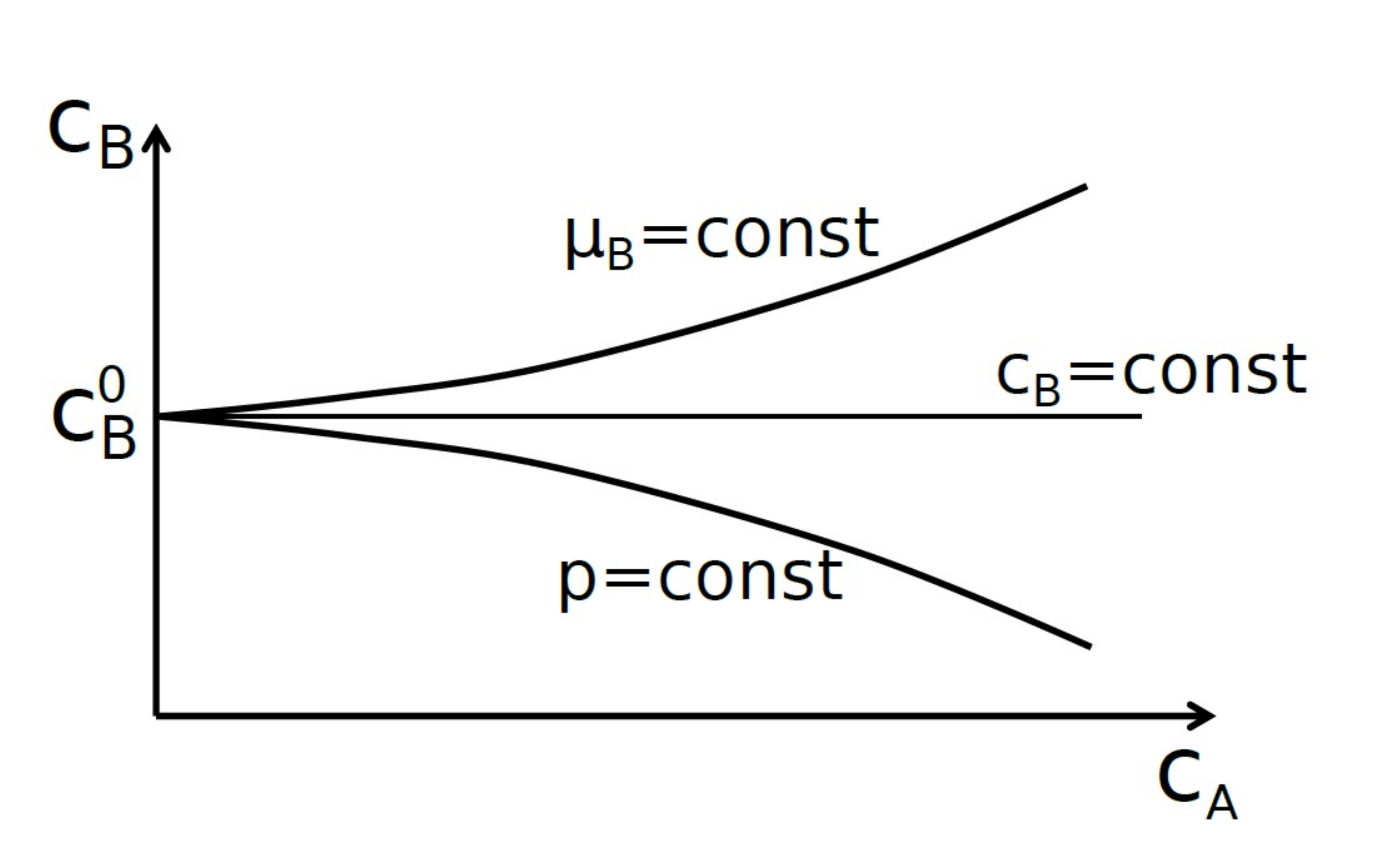}
\caption{The  intensive properties of a two-component solution are functions of three independent intensive variables. We fix the temperature $T$ and examine the parameter space formed by two other intensive variables $c_A$ and $c_B$, showing schematic contours of the two intensive properties $\mu_B$ and $p$.  Starting at the initial point (pure solvent, with density $c_B^0$) with $c_A = 0$, the virial series for property $\mu_A$    along the curve $\mu_B = const$ will be converted to one along the curve $p= const$.}
\label{fig:parameterspace}
\end{figure}

The corresponding virial series for the solvent chemical potential $\mu_B$ is found from the $\mu_A$ series and the thermodynamic relation (\ref{eq4.23}). Substituting  the virial series (\ref{eq4.17}) for $\beta \mu_A$ on the right side  of (\ref{eq4.23}) gives

%===========================================
% % 4.27
\begin{equation}
\label{eq4.27}
 \frac{\partial \beta \mu_B}{\partial c_A} = -\frac{1}{c_B} - \hat{B} \frac{c_A}{c_B}~,~~~(p,T~fixed)~.
\end{equation}
%==========================================
We now integrate (\ref{eq4.27}) with respect to $c_A$, bearing in mind we require terms in $\mu_B$ only up to $O(c_A^2)$. In terms of the independent variables $\{c_A,p,T\}$, we therefore expand $c_B(c_A,p,T)$ to first order in $c_A$,

%==========================================
% % 4.28
\begin{eqnarray}
\label{eq4.28}
c_B &\doteq& c_B^0 + \frac{\partial c_B}{\partial c_A}c_A~,~~~(p,T~fixed)~ ,   \nonumber\\
    &=& c_B^0(1 + \hat{b}c_A )~,
\end{eqnarray}
%===================================================
where

%================================================
% % 4.29
\begin{equation}
\label{eq4.29}
\hat{b} = \frac{1}{c_B^0}\left(\frac{\partial c_B}{\partial c_A}\right)_{p,T,c_A\rightarrow0}~.
\end{equation}
%==============================================
We use (\ref{eq4.28}) in the first $1/c_B$ term in (\ref{eq4.27}) and expand $(1 + \hat{b}c_A)^{-1}$ as $(1-\hat{b}c_A)$ to first order. We can use $1/c_B \doteq 1/c_B^0$ in the second term, as corrections would generate terms of higher order  in $c_A$ than we need. With these substitutions the integrals are elementary and give, to $O(c_A^2)$,

%=======================================================
% % 4.30
\begin{equation}
\label{eq4.30}
\beta \mu_B = \beta\mu_B^0 - \frac{c_A}{c_B^0}- \frac{1}{2}(\hat{B}-\hat{b})\frac{c_A^2}{c_B^0}~,
\end{equation}
%=====================================================
where $\mu_B^0$ is the pure solvent chemical potential. 

Some thermodynamic manipulations will turn the expression (\ref{eq4.29}) for $\hat{b}$ into something more familiar. With constant $T$ to be understood in all derivatives to follow, and  remembering that it is the limiting quantity with $c_A\rightarrow0$ that we need, we first re-write the derivative $(\partial c_B/\partial c_A)_p$ occuring in the definition (\ref{eq4.29}) of $\hat{b}$ using the anti-chain rule as

%============================================
% % 4.30a
\begin{equation}
\label{eq4.30a}
\left(\frac{\partial c_B}{\partial c_A}\right)_p = -\left(\frac{\partial c_B}{\partial p}\right)_{c_A} \left(\frac{\partial p}{\partial c_A}\right)_{c_B}~.
\end{equation}
%=========================================
The first partial derivative factor on the right side in (\ref{eq4.30a}) becomes a pure solvent quantitity in the limit $c_A\rightarrow0$, and from the definition of the isothermal compressibility is given by

%========================================
% % 4.31a
\begin{equation}
\label{eq4.31a}
\left(\frac{\partial c_B}{\partial p}\right)_{c_A} = c_B^0 \chi_B~,~~~(c_A\rightarrow0)~.
\end{equation}
%============================================
 The second partial derivative factor on the right side in (\ref{eq4.30a}) can be re-written using the rule  for relating a partial derivative at fixed $c_B$ to one at fixed $\mu_B$,

%=============================================================
% % 4.32a
\begin{equation}
\label{eq4.32a}
\left(\frac{\partial p}{\partial c_A}\right)_{c_B} = \left(\frac{\partial p}{\partial c_A}\right)_{\mu_B} + \left(\frac{\partial p}{\partial \mu_B}\right)_{c_A} \left(\frac{\partial \mu_B}{\partial c_A}\right)_{c_B}~.
\end{equation}
%======================================================
The first partial derivative on the right side of the last equation, in the limit $c_A\rightarrow0$, is $k_BT$, which we find from the pressure virial series (\ref{eq4.12}). The second partial derivative on the right side  of (\ref{eq4.32a}), in the limit $c_A\rightarrow0$, becomes a pure solvent quantity which can be found from (\ref{eq2.12}), or from the Gibb-Duhem relation (\ref{eq4.13}), to be

%==================================================
% % 4.33a
\begin{equation}
\label{eq4.33a}
\left(\frac{\partial p}{\partial \mu_B}\right)_{c_A} = c_B^0~,~~~(c_A\rightarrow0)~.
\end{equation}
%=========================================================
	 Again in the limit of infinite dilution, the third partial derivative on the right side in (\ref{eq4.32a}) can be derived using the Raoult relation (\ref{eq4.24}), and in Appendix B (see (\ref{eqC.15}) and discussion) is shown  to be

%===================================================
% % 4.34a
\begin{equation}
\label{eq4.34a}
\left(\frac{\partial \mu_B}{\partial c_A}\right)_{c_B} = -\frac{k_BT}{c_B^0}+ \frac{\overline{v}_A^{\infty}}{c_B^0 \chi_B}~,~~~(c_A\rightarrow0)~.
\end{equation}
%=================================================
Combining the last three relations  we find

%==================================================
% %4.35a
\begin{equation}
\label{eq4.35a}
\left(\frac{\partial p}{\partial c_A}\right)_{c_B} = \frac{\overline{v}_A^{\infty}}{\chi_B}~,~~~(c_A\rightarrow0)~.
\end{equation}
%=================================================
Substituting (\ref{eq4.35a}) and (\ref{eq4.31a}) into (\ref{eq4.30a}) we get 

%=========================================
 % % 4.36a
\begin{equation}
\label{eq4.36a}
\left(\frac{\partial c_B}{\partial c_A}\right)_p = - c_B^0\overline{v}_A^{\infty}~,~~~(c_A\rightarrow0)~,
\end{equation}
%=====================================================
and then from (\ref{eq4.29})

%===================================================
% % 4.37a
\begin{equation}
\label{eq4.37a}
\hat{b} = -\overline{v}_A^{\infty}~.
\end{equation}
%=====================================

Thus the coefficient  $\hat{B} - \hat{b}$ in (\ref{eq4.30}) is $\hat{B} + \overline{v}_A^{\infty}$, and since from (\ref{eq4.25}) we have $\hat{B}= 2\tilde{B} - \overline{v}_A^{\infty} + k_BT \chi_B$, we find  $\hat{B} - \hat{b} = 2\tilde{B} + k_BT \chi_B$. Substituting this last result into (\ref{eq4.30}) gives the virial series  for $\mu_B$   at constant pressure  expressed in terms of the original second  virial coefficient $\tilde{B}$,

%====================================================
% % 4.31
\begin{equation}
\label{eq4.31}
\beta \mu_B(c_A,p,T) = \beta \mu_B^0(p,T) - \frac{c_A}{c_B^0} - \left(\tilde{B}(p,T) + \frac{1}{2} k_BT \chi_B(p,T)\right) \frac{c_A^2}{c_B^0}~,
\end{equation}
%=====================================================
where $c_B^0 = c_B^0(p,T)$. We see that the first virial coefficient here, $-{1/c_B^0(p,T)}$, is trivial in the sense used earlier, so that the first-order change in the solvent chemical potential upon addition of solute, unlike the corresponding change in solute chemical potential, is a colligative property like the osmotic pressure (Raoult law). The second virial coefficient for $\beta \mu_B$, i.e., $-\left(\tilde{B}(p,T) + \frac{1}{2} k_BT \chi_B(p,T)\right)/c_B^0(p,T)$, will be positive at the  temperatures where $\tilde{B}(p,T)<0$, in the usual situations where $k_BT\chi_B/2$ is negligible.

From the  constant pressure virial series (\ref{eq4.26}) and (\ref{eq4.31}) for $\mu_A$ and $\mu_B$ we obtain the first two terms in the constant pressure virial series for the Gibbs free energy $G$ using (\ref{eq4.16}),

%======================================================
% % 4.32
\begin{eqnarray}
\label{eq4.32}
\beta G(N_A,N_B,p,T) = \beta G_B(N_B,p,T) + \beta G_A^{ideal} + \left( \tilde{A}(p,T) -1 \right)N_A 
\nonumber  \\
+ \left( \tilde{B}(p,T) + \frac{1}{2} k_BT \chi_B(p,T) \right)\frac{N_A^2}{V_B}~,
\end{eqnarray}
%=========================================================
where $V_B = V_B(N_B,p,T)$ is the pure solvent volume, and $\beta G_A^{ideal} = N_A \ln(c_A \Lambda_A^3) $ is the solute ideal gas Gibbs free energy, which will be different for polyatomic solutes. In deriving (\ref{eq4.32}) from (\ref{eq4.26}) and (\ref{eq4.31}) we express everything in terms of the independent variables $\{N_A,N_B,p,T\}$, and we retain terms in $N_A$ only of orders $N_A$ and $N_A^2$. Thus we write $c_A = N_A/V$, where $V = V(N_A,N_B,p,T)$ is expanded to first order in $N_A$ as

%====================================================
% % 4.32x
\begin{equation}
\label{eq4.32x}
V \doteq V_B + \overline{v}_A^{\infty} N_A~,
\end{equation}
%===========================================
where $\overline{v}_A^{\infty}$ is the infinite dilution limit ($N_A\rightarrow0$) of the solute partial molecular volume $\overline{v}_A = (\partial V/ \partial N_A)_{N_B,p,T}$. Using (\ref{eq4.32x}) we express $c_A$ to $O(N_A^2)$ as 

%====================================================
% % 4.32y
\begin{equation}
\label{eq4.32y}
c_A \doteq \frac{N_A}{V_B}\left(1 - \frac{\overline{v}_A^{\infty}}{V_B}N_A \right)~.
\end{equation}
%=================================================
From the last relation and $c_B^0 = N_B/V_B$ we readily find (\ref{eq4.32}) from (\ref{eq4.26}) and (\ref{eq4.31}).

From (\ref{eq4.32}) we obtain the virial series at constant pressure for the enthalpy $H$ using (\ref{eq4.15}),

%=======================================================
% %4.33
\begin{equation}
\label{eq4.33}
\beta H(N_A,N_B,p,T) = \beta H_B(N_B,p,T) + \beta H_A^{ideal}+A^H(p,T)N_A +B^H(p,T)\frac{N_A^2}{V_B}~,
\end{equation}
%===============================================
where $\beta H_A^{ideal} = (5/2)N_A $ is the solute ideal gas value and the first two virial coefficients in the enthalpy series are given by

%====================================================
% % 4.34
\begin{equation}
\label{eq4.34}
A^H(p,T)= -T \frac{\partial}{\partial T} \tilde{A}(p,T)~,
\end{equation}
%==========================================
% % 4.35
\begin{equation}
\label{eq4.35}
B^H(p,T)=\left(-T \frac{\partial}{\partial T} + T \alpha_B(p,T) \right) \left( \tilde{B}(p,T) + \frac{1}{2} k_BT \chi_B(p,T) \right) ~,
\end{equation}
%==========================================
where $\alpha_B = (1/V_B)(\partial V_B/ \partial T)_{N_B,p}$ is the pure solvent thermal expansion coefficient. Because $\beta H_A^{ideal}$ is linear in $N_A$ we could instead include a term $5/2$  in $A^H(p,T)$ but choose not to do so as a reminder that this term will change for polyatomic solutes, and because the solute ideal gas term does not contribute to the enthalpy of mixing, which is often the quantity of interest and which is determined by the last two terms in (\ref{eq4.33}). Equations (\ref{eq4.34}) and (\ref{eq4.35}) will be useful in analyzing  ITC experiments as discussed briefly in the next section.

From the two virial series for $G(N_A,N_B,p,T)$ and $H(N_A,N_B,p,T)$ we can get one for the entropy $S(N_A,N_B,p,T)$ using $S/k_B = \beta H - \beta G $.

%=============================================
%
%
%

%
%
%
%====================================================\begin{figure}[hbtp]

%==============================================================
%
%
%
\section{DISCUSSION AND CONCLUDING REMARKS} 
%
%
%
%================================================================

 A simple example of a solute pair PMF $w_{AA}(r)$ determined from computer simulation using a nonequilibrium work method\cite{nategholeslam-et-al} for benzene in water at temperature $T = 303K$ is shown in Fig.2. Benzene is a classic hydrophobe with very low solubility in water. The experimental value $\tilde{B} = -715 \pm15 \AA^3 $ for the osmotic second virial coefficient at $T=303 K$ was obtained from heroic vapor pressure measurements\cite{tucker-et-al} using the concentration dependence of the Henry law constant $K_H$. The negative sign found for $\tilde{B}$ is that expected for attractive hydrophobic interaction. The theoretical value we obtain from the expression (\ref{eq4.11}) and the data of Fig.2 is $\tilde{B} = -998 \pm 218 \AA^3$, in reasonable agreement with the experimental value; the parameters for the vacuum pair potentials of the  benzene and water models were taken from the literature (see references in the caption of Fig.2), and data for $\tilde{B}$ is not used in determining these parameters. The PMF for benzene-benzene in water and the  osmotic second virial coefficient have been determined by a number of other workers\cite{linse, jorgensen-severance, gao, chipot-et-al, villa-et-al} from various simulation algorithms and vacuum pair potential models and a variety of state conditions, and by Rossky and Friedman\cite{rossky-friedman} with a simplified model for the benzene-benzene PMF. There are also some calorimetry data\cite{hallen-et-al} for the benzene-water system and if enthalpy osmotic virial coefficients can be extracted from them the equations (\ref{eq4.34}) and (\ref{eq4.35}), together with simulation results over a range of temperature for the benzene excess chemical potential and the benzene-benzene PMF, will provide the basis for an analysis.

\begin{figure}
\centering
\includegraphics[scale=0.5]{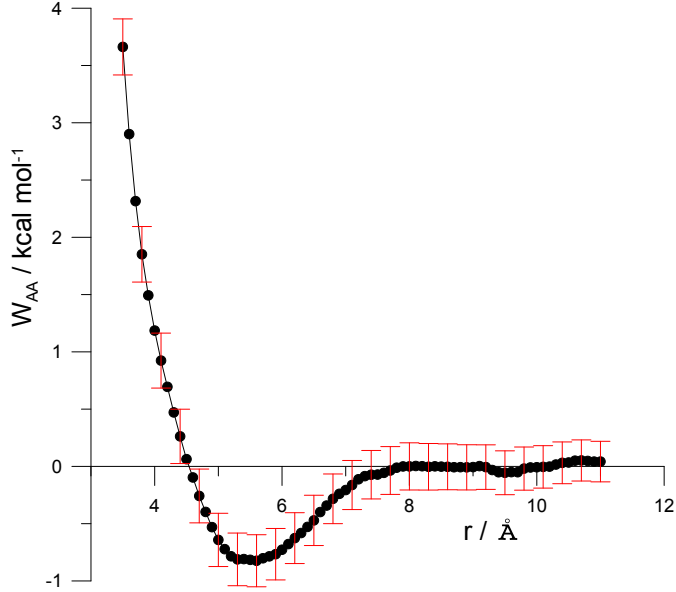}
\caption{PMF $w_{AA}(r)$ for benzene-benzene in water at T = 303K  obtained from MD simulation, with rigid molecular models for TIP3P water and for 12-site benzene constructed in VMD 1.9/Molefacture 1.3 environment. Force-field parameters from CHARMM27. PMF constructed from a 12ns $NVT$ simulation with 2fs time step using nonequilibrium work OFR algorithm\cite{nategholeslam-et-al} with OFR-AT analysis code\cite{holland-et-al} running on NAMD 2.8.}
\label{fig:benzene-pmf}
\end{figure}

We are currently analyzing ITC experimental data for the enthalpy of mixing of antimicrobial peptides at low concentrations in water \cite{nichols-et-al} and extracting enthalpy second virial coefficients. Using the new expression (\ref{eq4.35}) for the enthalpy second virial coefficient at constant pressure,  with the temperature derivative being obtained from simulation results for the peptide-peptide PMF $w_{AA}(r)$ taken at three temperatures, we will compare theory and experiment for this more complicated system in a future publication.

%%%%%%%%%%%%%%%%%%%%%%%%%%%%%%%%%%%%%%%%%%%%%%%%%%%%%%%%%%%%%%%%%%%%%%%%%%%%%%%%%%%%%%%%%%%%%%%%%%%%%%%%%%%%%%%%%%%%%%%%%%%%%
%%%%%%%%%%%%%%%%%%%%%%%%
\section*{Acknowledgements}
%%%%%%%%%%%%%%%%%%%%%%%%%%%%%%%%%%%%%%%%%%%%%%%%%%%%%%%%%%%%%%%%%%%%%%%%%%%%%%%%%%%%%%%%%%%%%%%%%%%%%%%%%%%%%%%%%%%%%%%%%%%%%
%%%%%%%%%%%%%%%%%%%%%%%%
%
We are grateful to NSERC for financial support, Compute Canada for computing support, and Mostsfa Nategholeslam for help with \LaTeX.
%%%%%%%%%%%%%%%%%%%%%%%%%%%%%%%%%%%%%%%%%%%%%%%%%%%%%%%%%%%%%%%%%%%%%%%%%%%%%%%%%%%%%%%%%%%%%%%%%%%%%%%%%%%%%%%%%%%%%%%%%%%%%
%%%%%%%%%%%%%%%%%%%%%%%

%==========================================================
%
%
%
\section*{Appendix A: Some details on the cluster decomposition of $W_A(\boldsymbol{r}_A^{N_A})$}
%
%
% 
%===================================================
We give a few more details on the derivation of the cluster decompostion (\ref{eq3.4}) of the total effective solute potential $W_A(\boldsymbol{r}_A^{N_A})$ up to the pairs term. In terms of the solvent-induced part $V_A$ of $W_A$ defined by (\ref{eq3.1}), we want to show

%===============================
% % A.1
\begin{align}
\label{eqA.1}
\langle \mbox{e}^{-\beta U_{AB}} \rangle_B~ &\equiv \mbox{e}^{- \beta V_A} \nonumber
 \\ &=  \mbox{e}^{-\beta(\sum_{i}v_{A_i} + \sum_{i<j}v_{A_iA_j} + ...)}~,
\end{align}
%================================================
where $v_{A_i}$ and $v_{A_iA_j}$ are defined by (\ref{eq3.6}) and (\ref{eq3.8}) respectively.

We write the total solute-solvent potential $U_{AB}$ as a sum of its contributions $U_{A_1B}, U_{A_2B} $, etc, from solute molecules $A_1, A_2 $, etc, so that

%========================================================
% % A.2
\begin{equation}
\label{eqA.2}
\langle\mbox{e}^{-\beta U_{AB}}
\rangle_B ~\equiv~ \langle \mbox{e}^{-\beta(U_{A_1B} + U_{A_2B}+...)}
\rangle_B~,
\end{equation}
%==========================================================
where the solute $A$-coordinates are fixed and the solvent $B$-coordinates are averaged. The solution is assumed dilute such that solute configurations with only singlets and pairs of $A$-molecules need be considered. As an example, consider an $A$-configuration where $A_1A_2$ are a close solute pair, $A_3A_4$ are a close pair, and the remaining solute molecules $A_5, A_6, A_7$, etc, are singlets. For this $A$-configuration we group the terms on the right side of (\ref{eqA.2}) accordingly as

%=======================================================
% % A.3
\begin{equation}
\label{eqA.3}
\langle \mbox{e}^{-\beta U_{AB}}\rangle{_B}~=~ \langle \mbox{e}^{-\beta (U_{A_1B}+ U_{A_2B})}~ \mbox{e}^{-\beta( U_{A_3B}+U_{A_4B})}~ \mbox{e}^{-\beta U_{A_5B}}~\mbox{e}^{-\beta U_{A_6B}}~...
\rangle_B~.  
\end{equation}
%===================================================
In the averaging in (\ref{eqA.3}) the exponential factors can be taken to be uncorrelated because of the assumed isolation of the indicated pairs and singlets so that we have

%================================================
% % A.4
\begin{equation}
\label{eqA.4}
\langle \mbox{e}^{-\beta U_{AB}}\rangle_B ~ = \langle \mbox{e}^{-\beta(U_{A_1B} + U_{A_2B})}\rangle_B  ~\langle \mbox{e}^{-\beta(U_{A_3B}+U_{A_4B})}\rangle  _B~ \langle \mbox{e}^{-\beta U_{A_5B}}\rangle_B ~ \langle \mbox{e}^{-\beta U_{A_6B}}\rangle  _B...~~.  
\end{equation}
%==============================================
When we examine the definitions of $v_{A_i}$ and $v_{A_iA_j}$ in (\ref{eq3.6}) and (\ref{eq3.8}), we see that in (\ref{eqA.4}) a factor of $\mbox{e}^{-\beta v_{A_i}}$ arises for every $A$-molecule $A_i$; the first average factor $\langle ...
\rangle_B$ on the right side of (\ref{eqA.4}) generates $\mbox{e}^{-\beta v_{A_1}} \times \mbox{e}^{-\beta v_{A_2}}$, the second average generates  $\mbox{e}^{-\beta v_{A_3}} \times\mbox{e}^{-\beta v_{A_4}}$, the next average generates $\mbox{e}^{-\beta v_{A_5}}$, the next  $\mbox{e}^{-\beta v_{A_6}}$, etc. A pairs average factor $\mbox{e}^{-\beta v_{A_iA_j}}$ arises from just the close pairs average factors in  (\ref{eqA.4}), i.e.,   $\mbox{e}^{-\beta v_{A_1A_2}}$ from the first average and  $\mbox{e}^{-\beta v_{A_3A_4}}$ from the second.

The general pattern should now be clear: the averaging in (\ref{eqA.2}) generates a factor of $\mbox{e}^{-\beta v_{A_i}}$ for every $A$-molecule and a factor of  $\mbox{e}^{-\beta v_{A_iA_j}}$ for every close pair of $A$-molecules. But because $v_{A_iA_j}$ vanishes for large separation $r_{ij}$, we can include a factor   $\mbox{e}^{-\beta v_{A_iA_j}}$ for every solute pair, and thus replace $\mbox{e}^{-\beta \sum_{i<j}^{close~pairs}v_{A_iA_j}}$ with $\mbox{e}^{-\beta \sum_{i<j}^{all~pairs}v_{A_iA_j}}$ . We thus see how (\ref{eqA.2}) has the form (\ref{eqA.1}).

%=========================================================
%
%
%
\section*{Appendix B: Some needed thermodynamic relations}
%
%
%
%========================================================

In Section IV.C we used some thermodynamic relations which we now derive. For the first one, equation (\ref{eq4.22}), we write the quantity needed $(\partial \mu_A/\partial \mu_B)_{c_A}$ using the chain rule as

%========================================
% % C.1
\begin{equation}
\label{eqC.1}
\left(\frac{\partial \mu_A}{\partial \mu_B}\right)_{c_A} = \left(\frac{\partial \mu_A}{\partial c_B}\right)_{c_A} \left(\frac{\partial c_B}{\partial \mu_B}\right)_{c_A}~,
\end{equation}
%=============================================
where here and in the rest of this appendix, if not indicated explicitly, all partial derivatives are understood to be  at fixed $T$, and we recall we need the various quantities in the dilute limit $c_A\rightarrow0$. 
For the first partial derivative on the right side of (\ref{eqC.1}) we use the reciprocal relation

%==============================================
% %C.1a
\begin{equation}
\label{eqC.1a}
\left(\frac{\partial \mu_A}{\partial c_B}\right)_{c_A} =\left(\frac{\partial \mu_B}{\partial c_A}\right)_{c_B}~,
\end{equation}
which  is readily derived from (\ref{eq2.11}). For $c_A\rightarrow0$ the derivative on the right side of (\ref{eqC.1a}) is shown below to be given by

%============================================
% % C.1b
\begin{equation}
\label{eqC.1b}
\left(\frac{\partial \mu_B}{\partial c_A}\right)_{c_B} = -\frac{k_BT}{c_B^0} + \frac{\overline{v}_A^{\infty}}{c_B^0 \chi_B}~,~~~(c_A \rightarrow0)~,
\end{equation}
where $\overline{v}_A^{\infty}$ is the solute partial molecular volume at infinite dilution, and $c_B^0$ and $\chi_B$ are the pure solvent density and isothermal compressibility, respectively.  The second partial derivative on the right side of (\ref{eqC.1}) becomes a pure solvent quantity in the limit $c_A \rightarrow0$ and using the chain rule $(\partial c/ \partial \mu) = (\partial c / \partial p )(\partial p/ \partial \mu)$, the definition of isothermal compressibility for $(\partial c / \partial p)$, and (\ref{eq4.33a}) for $(\partial p / \partial \mu)$, is readily shown to be

%=======================================
% % C.1c
\begin{equation}
\label{eqC.1c}
 \left(\frac{\partial c_B}{\partial \mu_B}\right)_{c_A} = \chi_B {c_B^0}^2~,~~~(c_A \rightarrow0)~.
\end{equation}
From the last four relations we thus find

%======================================= 
% % C.1d
\begin{equation}
\label{eqC.1d}
\left(\frac{\partial \mu_A}{\partial \mu_B}\right)_{c_A} = c_B^0(\overline{v}_A^{\infty} - k_BT \chi_B)~,~~~(c_A \rightarrow0)~,
\end{equation}
which is the relation (\ref{eq4.22}) being sought.

The next relation we need to derive is (\ref{eqC.1b}) above, which is also (\ref{eq4.34a}) of the text. This relation can be derived using statistical mechanics/potential distribution theory\cite{gregorio-widom}, and we give here an alternative derivation from thermodynamics.  The first term on the right side of (\ref{eqC.1b}) is the limiting value of $(\partial \mu_B/\partial c_A)_{p,T}$ as seen from the Raoult relation (\ref{eq4.24}). The general relation between the derivatives $(\partial\mu_B/\partial N_A)_V$ and $(\partial\mu_B/\partial N_A)_p$ is

%===========================================
% % C.10
\begin{equation}
\label{eqC.10}
\left(\frac{\partial \mu_B}{\partial N_A}\right)_{N_B,V,T} = \left(\frac{\partial \mu_B}{\partial N_A}\right)_{N_B,p,T} + \left(\frac{\partial \mu_B}{\partial p}\right)_{N_A,N_B,T} \left(\frac{\partial p}{\partial N_A}\right)_{N_B,V,T}~.
\end{equation}
%==========================================
Since $\mu_B$ is intensive, we can write the derivative on the left side  as $(1/V)(\partial \mu_B/\partial c_A)_{c_B,T}$. On the right side, using the chain rule  we write the first partial derivative as

%===============================================
% % C.11
\begin{equation}
\label{eqC.11}
\left(\frac{\partial \mu_B}{\partial N_A}\right)_{N_B,p,T} = \left(\frac{\partial \mu_B}{\partial c_A}\right)_{N_B,p,T} \left(\frac{\partial c_A}{\partial N_A}\right)_{N_B,p,T}~.
\end{equation}
%=====================================================
In this equation, because $\mu_B$ is intensive we can write the first factor on the right side as $(\partial \mu_B/\partial c_A)_{p,T}$, and using $c_A = N_A/V$ we find $(1-c_A \overline{v}_A)/V$ for the second factor on the right side, where $\overline{v}_A$ is the solute partial molecular volume, so that we have

%=================================================
% % C.12
\begin{equation}
\label{eqC.12}
\left(\frac{\partial \mu_B}{\partial N_A}\right)_{N_B,p,T} =  \left(\frac{\partial \mu_B}{\partial c_A}\right)_{p,T}\left(\frac{1-c_A \overline{v}_A}{V}\right)~.
\end{equation}
%=============================================
The second partial derivative on the right side of (\ref{eqC.10}) is shown below to have the value

%======================================
% %C.12a
\begin{equation}
\label{eqC.12a}
\left(\frac{\partial \mu_B}{\partial p }\right)_{N_A,N_B,T} = \overline{v}_B~,
\end{equation}
%=====================================
where $\overline{v}_B$ is the solvent partial molecular volume. We  re-write the last factor on the right side of (\ref{eqC.10}) using the anti-chain rule as

%==================================================
% % C.13
\begin{equation}
\label{eqC.13}
\left(\frac{\partial p}{\partial N_A}\right)_{N_B,V,T} =- \left(\frac{\partial p}{\partial V}\right)_{N_A,N_B,T}\left(\frac{\partial V}{\partial N_A}\right)_{N_B,p,T}~.
\end{equation}
%=================================================
On the right side of the last equation, the first factor is given by the definition $(\partial p/\partial V)_{N_A,N_B,T} \equiv -1/\chi V$ where $\chi$ is the solution isothermal compressibility, and the second factor is, by definition, the solute partial molecular volume $\overline{v}_A$, so that we have

%================================================
% % C.14
\begin{equation}
\label{eqC.14}
\left(\frac{\partial p}{\partial N_A}\right)_{N_B,V,T} = \frac{\overline{v}_A}{\chi V}~.
\end{equation}
%======================================================
Substituting the various derivatives we have found into (\ref{eqC.10}) and cancelling a common factor of $1/V$ gives the general relation

%====================================================
% % C.15
\begin{equation}
\label{eqC.15}
\left(\frac{\partial \mu_B}{\partial c_A}\right)_{c_B} = \left(\frac{\partial \mu_B}{\partial c_A}\right)_{p}(1-c_A \overline{v}_A) + \frac{\overline{v}_A \overline{v}_B}{\chi}~,
\end{equation}
%====================================================
where constant $T$ is to be understood in the two partial derivatives here. In this relation, when we take the limit $c_A\rightarrow0$ for an infinitely dilute solution, the left side is what we need in (\ref{eqC.1b}), the partial derivative on the right side becomes $-k_BT/c_B^0$ by the Raoult relation (\ref{eq4.24}), the factor $(1-c_A\overline{v}_A)$ becomes unity, and the last term becomes $\overline{v}_A^{\infty}/c_B^0 \chi_B$. This completes the derivation of (\ref{eqC.1b}) (and (\ref{eq4.34a})), subject to proving (\ref{eqC.12a}).

To establish the last relation we need, equation (\ref{eqC.12a}), we start with the basic thermodynamic relation for the differential of the Gibbs free energy $dG$,

%===================================
% % C.16
\begin{equation}
\label{eqC.16}
dG = -S dT + V dp +\mu_A dN_A + \mu_B dN_B~.
\end{equation}
%=====================================
Comparing coefficients of $dN_B$ and $dp$ and equating cross derivatives gives

%==============================================
% % C.17
\begin{equation}
\label{eqC.17}
\left( \frac{\partial \mu_B}{\partial p} \right)_{N_A,N_B,T} = \left( \frac{\partial V}{\partial N_B} \right)_{N_A,p,T}~.
\end{equation}
%==================================================
By definition, the derivative on the right side of the last equation is the solvent partial molecular volume $\overline{v}_B$, and this establishes (\ref{eqC.12a}).

\end{document}